\documentclass[11pt,a4paper]{amsart}
\usepackage{amsaddr}

\usepackage{centernot}
\usepackage{epsfig}
\usepackage{etoolbox} 
\usepackage{thmtools}
\usepackage{thm-restate}
\usepackage{amssymb} 
\usepackage{url}
\usepackage{arydshln}
\usepackage{psfrag} 
\usepackage{marvosym}
\usepackage{amsmath} 
\usepackage{amsthm}
\usepackage{hyperref}
\usepackage{cancel}
\usepackage{ushort} 
\usepackage{orcidlink}

\theoremstyle{plain}
\newtheorem{theorem}{Theorem}[section]
\theoremstyle{remark}
\newtheorem{remark}[theorem]{Remark}

\newcommand{\mc}[1]{\mathcal{#1}}
\newcommand{\Comment}[1]{}
\newcommand{\todo}[1]{{\color{red}{#1}}}
\newcommand{\il}[1]{{\color{red}{#1}}}
\newcommand{\iu}[1]{{\color{red}{#1}}}
\newcommand{\len}[1]{|#1|}

\newcommand{\tran}[1]{\xrightarrow[]{#1}} 
\newcommand{\rtran}[1]{\stackrel{#1}{\rightsquigarrow}} 

\newcommand{\ftran}[1]{\stackrel{#1}{\rightharpoonup}} 
\newcommand{\ptran}[1]{\xrightarrow[]{#1}_{*}} 
\newcommand{\sqeqt}{\sim}
\newcommand{\Proc}{\mathsf{Proc}}
\newcommand{\Lab}{\mathsf{Lab}}

\newcommand{\Irr}{\mathsf{Irr}}

\newcommand{\lab}{\ell}

\newcommand{\Par}{\mid}

\newcommand{\rev}[1]{\ushort{#1}} 
\newcommand{\revw}[1]{\ushortw{#1}} 

\newcommand{\ceqt}{\approx}

\newcommand{\cte}{\sharp}
\newcommand{\ind}{\mathrel{\iota}}
\newcommand{\notind}{\mathrel{\,\,\iota\!\!\!\diagup}}
\newcommand{\leqf}{\leq_{\mathsf{f}}}
\newcommand{\ltf}{<_{\mathsf{f}}}

\newcommand{\es}{\varepsilon}
\newcommand{\cell}[1]{\langle #1 \rangle}
\newcommand{\co}{\mathrel{\mathsf{co}}} 
\newcommand{\coind}{\mathrel{\mathsf{ci}}} 
\newcommand{\indt}{_{\ind}} 
\newcommand{\ci}{_{\mathsf{ci}}} 
\newcommand{\cf}{\mathrel{\#}} 

\newcommand{\nil}{\mathbf{0}}
\newcommand{\und}[1]{\mathsf{und}(#1)}
\newcommand{\op}{^{\dagger}} 

\def\finex{{\unskip\nobreak\hfil
\penalty50\hskip1em\null\nobreak\hfil$\diamond$
\parfillskip=0pt\finalhyphendemerits=0\endgraf}}

\newcommand{\bydef}{\stackrel{\emph{def}}{=}}

\newcommand{\preS}[1]{\ensuremath{{}^\bullet{#1}}}

\newtoggle{tech_report}

\toggletrue{tech_report}
	
\iftoggle{tech_report}{

}{
}

\makeatletter

\theoremstyle{plain}
\newtheorem{proposition}[theorem]{Proposition}
\newtheorem{lemma}[theorem]{Lemma}
\newtheorem{corollary}[theorem]{Corollary}

\theoremstyle{definition}
\newtheorem{definition}[theorem]{Definition}

\theoremstyle{remark}
\newtheorem{example}[theorem]{Example}

\begin{document}


\title{An Axiomatic Theory for Reversible Computation}

\author{Ivan Lanese\,
\orcidlink{0000-0003-2527-9995}}
  \address{Olas Team, University of Bologna/INRIA, Italy}
    \email{ivan.lanese@gmail.com}

\author{Iain Phillips\,
\orcidlink{0000-0001-5013-5876}}
  \address{Imperial College London, England}
  \email{i.phillips@imperial.ac.uk}

 \author{Irek Ulidowski\,
\orcidlink{0000-0002-3834-2036}}
  \address{University of Leicester, England}
  \address{AGH University of Science and Technology, Krak\'{o}w, Poland}
  \email{i.ulidowski@leicester.ac.uk}

\renewcommand{\shortauthors}{Lanese, Phillips and Ulidowski}


\begin{abstract}
Undoing computations of a concurrent system is beneficial in many
situations, e.g., in reversible debugging of multi-threaded programs
and in recovery from errors due to optimistic execution in parallel
discrete event simulation. A number of approaches have been proposed
for how to reverse formal models of concurrent computation including
process calculi such as CCS, languages like Erlang, and abstract
models such as prime event structures and occurrence nets.  However it
has not been settled what properties a reversible system should enjoy,
nor how the various properties that have been suggested, such as the
parabolic lemma and the causal-consistency property, are related.  We
contribute to a solution to these issues by using a generic labelled
transition system equipped with a relation capturing whether
transitions are independent to explore the implications between
various reversibility properties.  In particular, we show how
all properties we consider are derivable from a set of axioms.
Our intention is that when establishing properties of some formalism
it will be easier to verify the axioms rather than proving properties
such as the parabolic lemma directly.  We also introduce two new
properties related to causal consistent reversibility, namely
causal liveness and causal safety, stating, respectively, that an
action can be undone if (causal liveness) and only if (causal
  safety) it is independent from all the following
actions. These properties come in three flavours:
  defined in terms of independent transitions, independent events, or
  via an ordering on events.  Both causal liveness and causal safety
are derivable from our axioms.



\end{abstract}
%

\maketitle

\section{Introduction}
Reversible computing studies computations which can proceed both in
the standard, forward direction, and backward, going back to past
states. Reversible computation has attracted interest due to its
applications in areas as different as low-power computing~\cite{Landauer61},
simulation~\cite{CarothersPF99}, robotics~\cite{LaursenSE15},
biological modelling~\cite{CardelliL11,PhillipsUY12} and
debugging~\cite{microsoft,LaneseNPV18b}. 

There is widespread agreement in the literature about what properties
characterise reversible computation in the classical sequential (hence deterministic)
  a notion of reversibility 
    suited for concurrent systems 
    called
\emph{causal-consistent reversibility} (other notions were also used later on, e.g., to model biological systems~\cite{PhillipsUY12}). According to
an informal account of causal-consistent reversibility, any action can be undone provided that its 
consequences\footnote{By consequence we mean any subsequent transition which could not be permuted with $t$ while preserving the resulting state.}
if any, are undone beforehand.
Following~\cite{DK04} this account is formalised using
the notion of causal
equivalent traces: two traces are causal equivalent if and only if they only
differ for swapping independent actions, and inserting or removing
pairs of an action and its reverse. According to~\cite[Section~3]{DK04}
\begin{quote}
Backtracking an event is possible when and only when a causally
equivalent trace would have brought this event as the last one
\end{quote}
which is then formalised as the so called causal
consistency (CC)~\cite[Theorem 1]{DK04}, stating that coinitial computations are
causal equivalent if and only if they are cofinal.
Our new proof of CC (Proposition~\ref{prop:PL WF CC})
shows that it holds in essentially any reversible formalism satisfying the
Loop Lemma (roughly, any action can be undone) and the Parabolic Lemma (roughly, any computation is equivalent to a backward computation followed by a forward one),
and we believe that CC is insufficient on its own to capture the informal notion.

A formalisation closer to the informal statement above is provided
in~\cite[Corollary 22]{LaneseNPV18}, stating that a forward transition
$t$ can be undone after a derivation if and only if all the consequences of~$t$, if any,
are undone beforehand. We are not aware of other discussions trying to
formalise such a notion, except for~\cite{PU15}, in the
setting of reversible event structures.
In~\cite{PU15}, a reversible event structure is \emph{cause-respecting}
if an event cannot be reversed until all events it has caused have also been reversed;
it is \emph{causal} if it is cause-respecting and a reversible event can be reversed
if all events it has caused have been reversed~\cite[Definition~3.34]{PU15}.

We provide (Section~\ref{sec:CSCL}) a novel definition of the idea above, composed by:
\begin{description}
\item[Causal Safety (CS):]
an action cannot be reversed until any actions caused by it have been reversed;
\item[Causal Liveness (CL):]
we should allow actions to reverse in any order compatible with CS, not necessarily 
the exact inverse of the forward order.
\end{description}
We shall see that CC does not capture the same property as CS+CL
(Examples~\ref{ex:prerev not CSi},~\ref{ex:prerev not CL} and~\ref{ex:CS CL not CC}), and that there are slightly
different versions of CS and CL, which can all be proved under
a small set of reasonable assumptions.

The main aim of this paper is to take an abstract model, namely
labelled transition systems with independence equipped with reverse
transitions (Section~\ref{sec:LTSIs}), and to show that the properties
above (as well as others) can be derived from a small set of simple
axioms (Sections~\ref{sec:basic},~\ref{sec:events},~\ref{sec:CSCL} and~\ref{sec:coinitial}). This is in sharp contrast with the large part of
works in the literature, which consider specific frameworks such as
CCS~\cite{DK04}, CCS with broadcast~\cite{Mez18}, CCB~\cite{KU18},
$\pi$-calculus~\cite{CristescuKV13}, higher-order
$\pi$~\cite{LaneseMS16}, Klaim~\cite{GiachinoLMT17}, Petri
nets~\cite{MMU19}, $\mu$Oz~\cite{LienhardtLMS12} and
Erlang~\cite{LaneseNPV18}, and all give similar but formally unrelated
proofs of the same main results. Such proofs will become instances of
our general results.
%
%
More precisely, our axioms will:
\begin{itemize}
\item
exclude behaviours which are not compatible with causal-consistent
reversibility (as we will discuss shortly);
\item
allow us to derive the main properties of reversible calculi which have been studied in the literature, 
such as CC (Proposition~\ref{prop:PL WF CC});
\item
hold for a number of reversible calculi which have been proposed, such
as RCCS~\cite{DK04} and reversible Erlang~\cite{LaneseNPV18} (Section~\ref{sec:casestudies}).
\end{itemize}
Thus, when defining a new reversible formalism, one just has to check
whether the axioms hold, and get for free the proofs of the most
relevant properties. Notably, the axioms are normally easier to prove
than the properties, hence the assessment of a reversible calculus
gets much simpler.

As a reference, Table~\ref{t:list} lists the axioms and properties used in this paper.
{\small 
\begin{table}[t!]
  \begin{center}
    \begin{tabular}{|c|c|c|c|c|} 
      \hline
      \textbf{Acronym} & \textbf{Name} & \textbf{Defined in} & \textbf{Proved in} & \textbf{Using}\\
      \hline
      SP & Square Property & Def.~\ref{def:basic} & Axiom & -\\
      BTI & Backward Transitions are Independent & Def.~\ref{def:basic} & Axiom & - \\
      WF & Well-Founded & Def.~\ref{def:basic} & Axiom & -\\
      PCI & Propagation of Coinitial Independence & Def.~\ref{def:PCI} & Axiom & implied by LG or CLG \\ \hdashline
      IRE & Independence Respects Events & Def.~\ref{def:IRE} & Axiom & implied by LG \\
      CIRE & Coinitial Independence Respects Events & Def.~\ref{def:CIRE} & Axiom & implied by IRE or CLG \\
      BFCIRE & Backward-Forward CIRE & Def.~\ref{def:BFCIRE} & Axiom & implied by CIRE\\

      IEC & Independence of Events is Coinitial & Def.~\ref{def:IEC} & Axiom & - \\ \hdashline
      CLG & Coinitial Label-Generated & Def.~\ref{def:CLG} & Str.\ Ax. & - \\
      LG & Label-Generated & Def.~\ref{def:LG} & Str.\ Ax. & - \\
      IC & Independence is Coinitial & Def.~\ref{def:coinitial LTSI} & Str.\ Ax. & implied by CLG \\
      \hline
      PL & Parabolic Lemma & Def.~\ref{def:PL} & Prop.~\ref{prop:PL} & BTI, SP\\
      CC & Causal Consistency & Def.~\ref{def:cc} & Prop.~\ref{prop:PL WF CC} & WF, PL\\
      UT & Unique Transition & Def.~\ref{def:ut} & Cor.~\ref{cor:ut} & CC\\
      BLD & Backward Label Determinism & Def.~\ref{def:BD} & Prop.~\ref{prop:BD}  & SP, BTI, PCI\\ 
      ID & Independence of Diamonds & Def.~\ref{def:ID} & Prop.~\ref{prop:ID} & BTI, PCI\\
      NRE & No Repeated Events & Def.~\ref{def:NRE} & Prop.~\ref{prop:NRE}  & Pre-rev.\\ 
      RPI & Reversing Preserves Independence & Def.~\ref{def:rpi} & Prop.~\ref{prop:RPI}  & SP, PCI, IRE, IEC\\ 
      CS$\indt$ & Causal Safety & Def.~\ref{def:safe live} & Thm.~\ref{thm:CS} & Pre-rev., IRE\\ 
      CL$\indt$ & Causal Liveness & Def.~\ref{def:safe live} & Thm.~\ref{thm:CL} & Pre-rev., IRE\\ 
      ECh & Event Coherence & Def.~\ref{def:ECh} & Prop.~\ref{prop:ECh} & Pre-rev., (IRE or IEC) \\
      CS$\ci$ & coinitial Causal Safety & Def.~\ref{def:coind safe live} & Thm.~\ref{thm:CS coind} & Pre-rev.\\ 
      CL$\ci$ & coinitial Causal Liveness & Def.~\ref{def:coind safe live} & Thm.~\ref{thm:CL coind} & Pre-rev., BFCIRE\\ 
      CS$_<$ & ordered Causal Safety & Def.~\ref{def:safe live <} & Prop.~\ref{prop:CS CL coind <} & Pre-rev.\\ 
      CL$_<$ & ordered Causal Liveness & Def.~\ref{def:safe live <} & Prop.~\ref{prop:CS CL coind <} & Pre-rev., BFCIRE\\ 
      \hline
    \end{tabular}
  \end{center}
    \caption{Axioms and properties for causal reversibility.
`Str.\ Ax.' abbreviates `Structural Axiom' and `Pre-rev.'  abbreviates `Pre-reversible', namely SP, BTI, WF, PCI (cf.~Def.~\ref{def:prerev}). We call statements in the bottom part of the table, namely from PL to CL$_<$, properties. }
    \label{t:list}  
\end{table}
}


In order to understand which kinds of behaviours are incompatible with
a causal-consistent reversible setting, consider the following CCS processes and their transitions as in Figure~\ref{fig:ccs1}:
\begin{figure}[!t]
\psfrag{a}{$a$}
\psfrag{b}{$b$}
\psfrag{0}{$\nil$}
\psfrag{a.0}{$a.\nil$}
\psfrag{b.0}{$b.\nil$}
\psfrag{a.0+b.0}{$a.\nil+b.\nil$}
\psfrag{P}{$P=a.P$}
\begin{center}
\epsfig{file=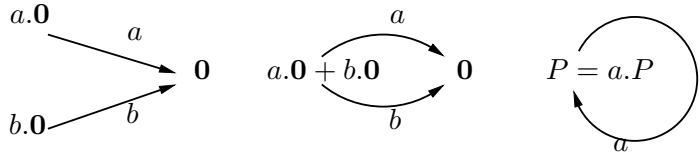, height=2.0cm}
\caption{Irreversible transition systems in CCS.
}\label{fig:ccs1}
\end{center}
\end{figure}

\begin{description}
\item[$a.\nil \tran a \nil$, $b.\nil \tran b \nil$:] from state $\nil$ one does not know whether to go back to $a.\nil$ or to $b.\nil$;
\item[$a.\nil + b.\nil \tran a \nil$, $a.\nil + b.\nil \tran b \nil$:] as above, but starting from the same process, hence showing that it is not enough to remember the initial configuration;
\item[$P \tran a P$ where $P \bydef a.P$:]  in state $P$ one does not know whether action $a$ has been performed, and, if 
it has been performed, then how many times. Due to this lack of information, one could go back an arbitrary number of times from 
$P$ to $P$.
In this work, we do not permit an arbitrary number of backward moves, because it goes against the idea that a state models a process reachable after a finite computation.

\Comment{
\iu{from state $P$ one can go back an arbitrary number of times to $P$}, which is against the idea that a state models a process reachable after a finite computation.
\todo{BLEND}
\il{in state $P$ one has no information about whether action $a$ has been performed, and in case how many times, hence one does not know whether it makes sense to undo $a$.}
}  
\end{description}
We remark that all such behaviours are perfectly reasonable in CCS,
and they are dealt with in the reversible setting by adding history information
about past actions. For example, in the first case one could remember the
initial state, in the second case both the initial state and the
action taken, and in the last case the number of iterations that have
been performed.

The paper is organised as follows. The next section introduces labelled transition systems with
independence (LTSIs). Three basic axioms for reversibility (SP, BTI and WF) are defined in Section~\ref{sec:basic}, 
and are used to prove the Parabolic Lemma and Causal Consistency.
Events are defined in Section~\ref{sec:events},
where another basic axiom (PCI) is formulated.
In Section~\ref{sec:CSCL} we discuss and define CS
and CL properties, and introduce three further basic axioms (IRE, CIRE and IEC)
that are used to prove them.
We consider three versions of CS and CL: those based on independence 
of transitions, on independence of events, and on ordering of events, and we study their relationships. 
We also show that axioms SP, BTI, WF, PCI and IEC, together with any one of IRE, CIRE and BFCIRE, are independent of each other.
Section~\ref{sec:coinitial} considers two structured forms of independence, namely independence defined on coinitial transitions only, and independence defined on labels only.
Eight case studies of reversible formalisms are presented in Section~\ref{sec:casestudies}, where we
demonstrate that our basic axioms are very effective in proving the main reversibility properties.
Section~\ref{sec:related} discusses relations with other works in the literature.
The final section contains concluding remarks and suggests potential future work.

This paper is an extended version of~\cite{LanesePU20}. The paper has been fully restructured, and now includes a number of additional or refined results. Beyond this, it includes full proofs of our results, as well as additional case studies, examples and explanations. We remark that the preliminary results in~\cite{LanesePU20} have already been exploited in~\cite{LaneseM20,AubertM21,BocchiLMY22,LamiLSCF22,BAGOSSY20223,Aub22,BM22}, which can be seen as further case studies for our approach.

\section{Labelled Transition Systems with Independence}\label{sec:LTSIs}
We want to study reversibility in a setting as general as possible.
Thus, we adopt initially only the core of the notion of \emph{labelled
 transition system with independence} (LTSI)~\cite[Definition~3.7]{SNW96},
and explore what can be achieved by adding 
various axioms on the independence relation. This is in contrast with the approach
taken in ~\cite{SNW96}, which requires a fixed number of axioms to hold in LTSIs.
Also, we extend LTSIs with reverse transitions, since we
study reversible systems. We first define labelled transition
systems (LTSs).

We consider the LTS of the entire set of processes in a calculus,
rather than the transition graph of a particular process and its derivatives, hence we do not fix an initial state.

\begin{definition}\label{def:lts}
A {\em labelled transition system (LTS)\/} 
is a structure \mbox{$(\Proc,\Lab,\tran{})$}, where $\Proc$ is the set of states (or processes),
$\Lab$ is the set of action labels and ${\tran{}}\subseteq \Proc \times \Lab \times \Proc$
is a transition relation. 
\end{definition}
We let $P,Q,\ldots$ range over processes,
$a,b,c,\ldots$ range over labels,
and $t,u,v,\ldots$ range over transitions (namely, elements of the transition relation).
We can write $t:P \tran a Q$ to denote that $t = (P,a,Q)$.
We call $a$-transition a transition with label $a$.
    
\begin{definition}[LTS with independence]\label{def:ltsi}
We say that $(\Proc,\Lab,\tran{},\ind)$ is an \emph{LTS with
  independence} (LTSI) if $(\Proc,\Lab,\tran{})$ is an LTS and $\ind$
is an irreflexive symmetric binary relation on transitions.
\end{definition}
In many cases (see Section~\ref{sec:casestudies}), the notion of
independence coincides with the notion of concurrency. However, this
is not always the case. Indeed, concurrency implies that transitions
are independent since they happen in different processses, but
transitions taken by the same process can be independent as
well. Think, for instance, of a reactive process that may react in any
order to two events arriving at the same time, and the final result
does not depend on the order of reactions.


We shall assume that all transitions are reversible,
so that the Loop Lemma~\cite[Lemma 6]{DK04} holds.
This does not hold in models of reversibility with control mechanisms~\cite{LaneseMS12}
such as irreversible actions~\cite{DK05} or a rollback operator~\cite{LMSS11}.
Nevertheless,
when showing properties of models with controlled reversibility it has
proved sensible to first consider the underlying models where all transitions
are reversible, and then study how control mechanisms change the
picture~\cite{GiachinoLMT17,LaneseNPV18}.
The present work helps with the first step.
\begin{definition}[Reverse and combined LTS]\label{def:reverse transition}
Given an LTS 
$(\Proc,\Lab,\ftran{})$,
let the \emph{reverse LTS} be $(\Proc,\Lab,\rtran{})$,
where $P \rtran a Q$ iff $Q \ftran a P$.
It is convenient to combine the two LTSs (forward and reverse):
let the reverse labels be
$\revw\Lab = \{\rev a : a \in \Lab\}$,
and define the \emph{combined LTS} to be
${\tran{}}\subseteq \Proc \times (\Lab \cup \revw\Lab) \times \Proc$
by $P \tran a Q$ iff $P \ftran a Q$ and $P \tran{\rev a} Q$ iff $P \rtran a Q$.
\end{definition}
We stipulate that the union $\Lab \cup \revw\Lab$ is disjoint.
We let $\alpha, \beta, \ldots$ range over $\Lab \cup \revw\Lab$.
For $\alpha \in \Lab \cup \revw\Lab$, the \emph{underlying} action label
$\und\alpha$ is defined as
$\und{a} = a$ and $\und{\rev a} = a$.
%
Let $\rev{\rev a} = a$ for $a \in \Lab$.  Given $t:P \tran \alpha Q$, let $\rev t:Q \tran {\rev\alpha} P$ be the transition which reverses $t$.
We define a labelling function $\lab$ from transitions to $\Lab\cup \revw\Lab$ by setting
$\lab((P,\alpha,Q)) = \alpha$.

We let $\rho,\sigma,\ldots$ range over finite sequences $\alpha_1 \ldots \alpha_n$,
with $\es$ representing the empty sequence. 
A \emph{path} is a sequence of forward or reverse transitions
of the form
$P_0 \tran{\alpha_1} P_1 \cdots \tran{\alpha_n} P_n$.
We let $r,s,\ldots$ range over paths.
We may write $r:P \ptran \rho Q$ where the intermediate states are understood.
On occasion we may refer to a path simply by its sequence of labels $\rho$.
The concatenation of paths $r$ and $s$ is written $rs$.
Given a path $r:P \ptran \rho Q$, the inverse path is $\rev r:Q \ptran {\rev \rho} P$
where $\rev\es = \es$ and $\rev{\alpha\rho} = \rev \rho \; \rev \alpha$.
The length of a path $r$ (notated $\len r$) is the number of transitions in the path.
Paths $P \ptran \rho Q$ and $R \ptran \sigma S$ are
\emph{coinitial} if $P = R$ and \emph{cofinal} if $Q = S$.
We say that a path is \emph{forward-only} if it contains no reverse transitions;
similarly a path is \emph{backward-only} if it contains no forward transitions.
Sometimes we let $f,\ldots$ and $b,\ldots$ range over forward-only and backward-only paths, respectively;
it will be clear from the context whether $b$ represents an action label or a path.

The irreversible processes in an LTS $(\Proc,\Lab,\tran{})$ are
$\Irr = \{P \in \Proc: P \not \rtran{}\}$.
A \emph{rooted path} is a
path $P \ptran \rho Q$ such that $P \in \Irr$.

In the following we consider LTSIs obtained by adding a notion of
independence to combined LTSs as above. We call the result a
\emph{combined LTSI}.

\begin{remark}
From now on, unless stated otherwise, we consider a combined LTSI $(\Proc,\Lab,\tran{},\ind)$. We will refer to it simply as an LTSI.     
\end{remark}

\section{Basic Properties}\label{sec:basic}
In this section we show that most of the properties in the reversibility literature (see, e.g.,
\cite{DK04,PU07,LaneseMS16,LaneseNPV18}), in particular the Parabolic Lemma and
Causal Consistency, can be proved under minimal assumptions on the
combined LTSI under analysis.

We formalise the minimal assumptions using three axioms, described below.
\begin{definition}[Basic axioms]\label{def:basic}
  We say an LTSI 
  satisfies:
  \begin{description}
\item[Square property (SP)]\!\!: if whenever $t:P \tran \alpha Q$, $u:P
    \tran \beta R$ with $t \ind u$ then there are cofinal transitions
    $u': Q \tran \beta S$ and $t':R \tran \alpha S$;
\item[Backward transitions are independent (BTI)]\!\!: if whenever $t:P \rtran{a} Q$ and $t': P \rtran{b} Q'$ 
     and $t \neq t'$ then $t \ind t'$;
\item[Well-founded (WF)]\!\!: if there is no infinite reverse
    computation, i.e.\ we do not have $P_i$ (not necessarily distinct)
    such that $P_{i+1} \tran {a_i} P_i$ for all $i = 0,1,\ldots$.
  \end{description}
\end{definition}
WF can alternatively be formulated using backward transitions,
but the current formulation makes
sense also in non-reversible calculi (e.g., CCS), which can be used as a comparison.
Let us discuss the intuition behind these axioms. SP takes its
name from the Square Lemma, where it is proved for concrete calculi and
languages~in~\cite{DK04,LaneseMS16,LaneseNPV18}, and
captures the idea that independent transitions can be executed in any
order, that is they form commuting diamonds. SP can be seen as a
sanity check on the chosen notion of independence. BTI
generalises the key notion of backward determinism used in sequential
reversibility (see, e.g., \cite{Pin87} for finite state
automata and \cite{YokoyamaG07} for the imperative
language Janus) to a concurrent setting.  Backward determinism can be
spelled as ``two coinitial backward transitions do coincide''. This can be generalised to
``two coinitial backward transitions
are independent''.
We will show in Proposition~\ref{prop:sequential} that the two definitions are equivalent  when no transitions are independent, 
which is the common setting in sequential computing.
Note that BTI and SP together rule out examples $a.\nil \tran a \nil$, $b.\nil \tran b \nil$ as well as $a.\nil + b.\nil \tran a \nil$, $a.\nil + b.\nil \tran b \nil$ from the Introduction.
Finally, WF means that we consider systems which have a
finite past. That is, we consider systems starting from some initial
state and then moving forward and back.
WF rules out example $P \tran a P$ where $P = a.P$ from the Introduction.

Axioms SP and BTI are related to properties which are part of the definition
of (occurrence) transition systems with independence in~\cite[Definitions~3.7, 4.1]{SNW96}.
WF was used as an axiom in~\cite{PU07a}.

%
%
%

Using the minimal assumptions above we can prove relevant results
from the literature. As a preliminary step, we define causal equivalence,
equating computations differing only for swaps of independent
transitions and simplification of a transition
with its reverse.

\begin{definition}[Causal equivalence, cf.~{\cite[Definition 9]{DK04}}]\label{def:ceqt}
  Consider an LTSI satisfying SP.
Let $\ceqt$ be the smallest equivalence relation on paths closed under composition and satisfying:
\begin{enumerate}
\item
(swap)  
if $t:P \tran \alpha Q$, $u:P \tran \beta R$ are independent,
and $u': Q \tran \beta S$, $t':R \tran \alpha S$ (which exist by SP)
then $tu' \ceqt ut'$;
\item
(cancellation)
$t \rev t \ceqt \es$ \quad and \quad $\rev t t \ceqt \es$.
\end{enumerate}
\end{definition}

We first consider the Parabolic Lemma \cite[Lemma
  10]{DK04}, which states that each path is causal equivalent to a
backward path followed by a forward path.

\begin{definition}\label{def:PL}
  {\bf Parabolic Lemma property (PL)}: for any path $r$ there are forward-only paths $s,s'$ such that 
$r \ceqt \rev s s'$ and $\len s + \len {s'} \leq \len r$.
\end{definition}

\begin{proposition}\label{prop:PL}
Suppose an LTSI satisfies BTI and SP.  Then PL holds.
\end{proposition}

\begin{proof}
Suppose BTI and SP hold.
Define a function on paths as follows:
$d(r)$ is the number of pairs of forward transitions $(t,u)$
such that $t$ occurs in any position to the left of $\rev u$ in $r$.
We say $r$ is parabolic iff $d(r) = 0$. We have to show that each path is causal equivalent to a parabolic one.

Suppose $d(r) > 0$.
We show that there is $s \ceqt r$ with $\len {s} \leq \len r$ and $d(s) < d(r)$.
Since $d(r) > 0$, we have $r = s_1 t \rev u s_2$ with
$s_1:P \tran {\sigma_1} R$,
$t:R \tran a S$, $\rev u:S \tran {\rev b} T$ and $s_2:T \tran {\sigma_2} Q$.
If $t = u$, then we obtain $r = s_1 t \rev u s_2 \ceqt s_1 s_2$.
Clearly $r \ceqt s_1s_2$ with $\len {s_1s_2} < \len r$ and $d(s_1s_2) < d(r)$.
So suppose $t \neq u$.
By BTI we have $\rev t \ind \rev u$.
By SP there are $S'$ and transitions $u':S' \tran b R$, $t':S' \tran a T$.
See Figure~\ref{fig:pl}.
\begin{figure}[!t]
\psfrag{a}{$a$}
\psfrag{b}{$b$}
\psfrag{s1}{$\sigma_1$}
\psfrag{s2}{$\sigma_2$}
\psfrag{P}{$P$}
\psfrag{Q}{$Q$}
\psfrag{R}{$R$}
\psfrag{S}{$S$}
\psfrag{T}{$T$}
\psfrag{S'}{$S'$}
\begin{center}
\epsfig{file=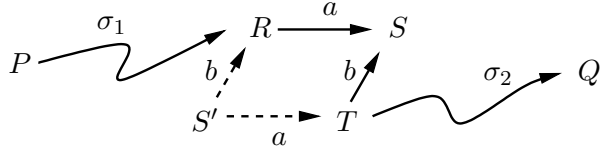, height=2.0cm}
\caption{Proof of Proposition~\ref{prop:PL}, case $t \neq u$.
}\label{fig:pl}
\end{center}
\end{figure}
Then $\rev t \, \rev{u'} \ceqt \rev u \, \rev{t'}$.
Hence, $r = s_1 t \rev u s_2
\ceqt s_1 t \rev u \, \rev{t'} t' s_2
\ceqt s_1 t \rev t \, \rev{u'} t' s_2
\ceqt s_1 \rev{u'} t' s_2 = s$ as required.
Given that $\len {s_1 \rev {u'} t' s_2} = \len r$ and $d(s_1 \rev {u'} t' s_2) = d(r)-1$ the thesis follows.
\end{proof}
The proof of Proposition~\ref{prop:PL} is very similar to that
of~\cite[Lemma~10]{DK04} except that in the latter BTI is shown directly as part of the proof.

A corollary of PL is that if a process is reachable
from an irreversible process, then it is also forwards reachable from it. In other words,
making a system reversible does not introduce new reachable states
but only allows one to explore forwards-reachable states in a different order. 
This is relevant, e.g., in reversible debugging of concurrent
systems~\cite{GiachinoLM14,LaneseNPV18}, where one wants to find bugs that actually 
occur in forward-only computations.
\begin{corollary}\label{freachable}
  Suppose an LTSI satisfies PL. If a process $P$ is reachable
  from some irreversible process $Q$, then it is also forward
  reachable from $Q$.
\end{corollary}
\begin{proof}
  By hypothesis, there is some path $r: Q \ptran{} P$. Thanks to PL,
  there are forward-only paths $s,s'$ such that $\rev s s': Q \ptran{}
  P$.  Since $Q$ is irreversible, $s = \es$, hence $s': Q \ptran{}
  P$ as desired.
\end{proof}
When WF and PL hold, each process is reachable from a unique irreversible process.
\begin{proposition}\label{prop:unique irrev}
Suppose an LTSI satisfies WF and PL.
For any process $P$ there is a unique irreversible process $I$ such that $P$ is reachable from $I$.
\end{proposition}
\begin{proof}
Let $P$ be any process.
We use WF to deduce that there is an irreversible process $I$ such that $P$ is (forward) reachable from $I$ via some path $r$.
Suppose now that $I'$ is irreversible  and there is a path $r'$ from $I'$ to $P$.
Then $r'\rev r: I' \ptran{} I$.
By PL there are forward-only paths $s,s'$ such that $\rev s s': I' \ptran{} I$.
But since $I$ and $I'$ are irreversible, both $s = \es$ and $s' = \es$.
Hence $I' = I$ as required.
\end{proof}

We now move to causal consistency~\cite[Theorem 1]{DK04}.

\begin{definition}\label{def:cc}
{\bf Causal Consistency (CC)}: if $r$ and $s$ are coinitial and cofinal paths then $r \ceqt s$.
\end{definition}

Essentially, causal consistency states that history information allows
one to distinguish computations which are not causal equivalent.
Indeed, if two computations are cofinal, that is they reach the same
final state (which includes the stored history information) then they
need to be causal equivalent.

Causal consistency frequently includes the other direction, namely that
coinitial causal equivalent computations are cofinal, meaning that
there is no way to distinguish causal equivalent computations. This
second direction follows easily from the definition of causal equivalence.

Notably, our proof of CC below is very much shorter than existing
proofs, such as the one of \cite[Theorem 1]{DK04} for RCCS and the one
of \cite[Theorem 21]{LaneseNPV18} for reversible Erlang.

\begin{proposition}\label{prop:PL WF CC}
Suppose an LTSI satisfies WF and PL. 
Then CC holds.
\end{proposition}
\begin{proof}
Let $r:P \ptran \rho Q$ and $r':P \ptran {\rho'} Q$.
Using WF, let $I,s$ be such that $s:I \ptran \sigma P$, $I \in \Irr$.
Now $sr\rev{sr'}$ is a path from $I$ to $I$,
and so by PL there are $r_1,r_2$ forward-only such that $\rev{r_1}r_2 \ceqt sr\rev{sr'}$.
But $I \in \Irr$ and so $r_1 = \es$ and $r_2 = \es$.
Thus $\es \ceqt sr\rev{sr'}$, so that $sr \ceqt sr'$ and (by composing with \rev{s} on the left) $r \ceqt r'$
as required.
\end{proof}

Causal equivalent computations are strongly related
in terms of the number
of transitions with a given label they contain.

\begin{proposition}\label{prop:count ceqt}
If $r \ceqt s$ then for any action $a$ the number of $a$-transitions in $r$ is the same as in $s$, where we count reverse transitions negatively.
\end{proposition}
\begin{proof}
  Straightforward, by induction on the derivation of $r \ceqt s$.
\end{proof}
\begin{remark}\label{rem:count fwd ceqt}
One consequence of 
Proposition~\ref{prop:count ceqt} is that if $r \ceqt s$ and $r$ and $s$
are both forward-only, then $\len r = \len s$.
\end{remark}

Causal consistency implies the unique transition property.

\begin{definition}\label{def:ut}
  \textbf{Unique transition (UT)}:
  if either $P \tran a Q$ and $P \tran b Q$ or $P \rtran a Q$ and $P \rtran b Q$ then $a = b$.
\end{definition}

\begin{figure}[!t]
\psfrag{BTI}{BTI}
\psfrag{WF}{WF}
\psfrag{PL}{PL}
\psfrag{CC}{CC}
\psfrag{BTI+WF}{BTI+WF}
\psfrag{BTI+CC}{BTI+CC}
\psfrag{PL+CC}{PL+CC}
\psfrag{PL+UT}{PL+UT}
\psfrag{WF+PL}{WF+PL}
\psfrag{WF+CC}{WF+CC}
\psfrag{prop:PL}{Prop.~\ref{prop:PL}}
\psfrag{prop:WFPLCC}{Prop.~\ref{prop:PL WF CC}}
\psfrag{ex:PLnotCC}{Ex.~\ref{ex:PL not CC}}
\psfrag{ex:notPL}{Ex.~\ref{ex:notPL}}
\psfrag{ex:notBTI}{Ex.~\ref{ex:notBTI}}
\psfrag{ex:notWF}{Ex.~\ref{ex:notWF}}
\psfrag{ex:WFnotCC}{Ex.~\ref{ex:WFnotCC}}
\begin{center}
\epsfig{file=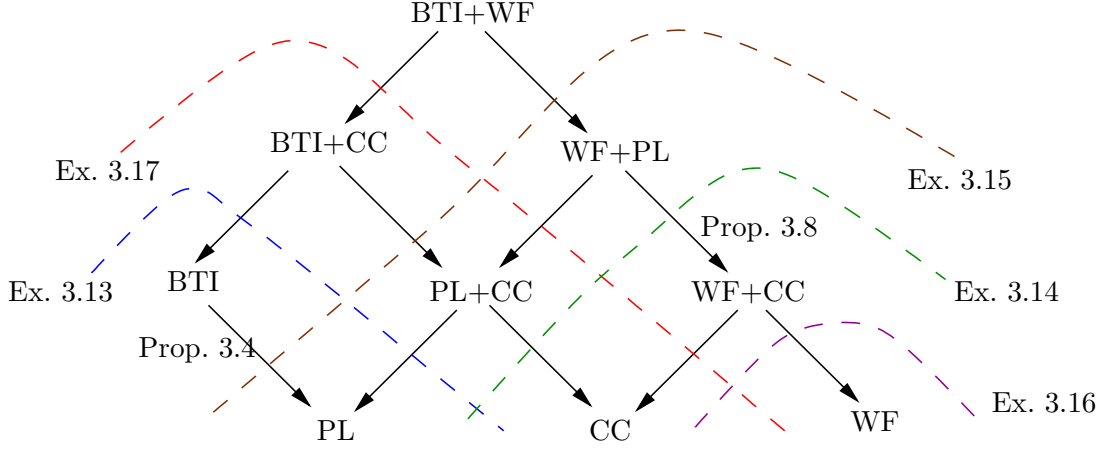, height=6cm}
\caption{Implications between the main properties discussed in Section~\ref{sec:basic}.
We assume SP throughout.
}\label{fig:basic2}
\end{center}
\end{figure}

\begin{corollary}\label{cor:ut}
  If an LTSI satisfies CC then it satisfies UT.
\end{corollary}
\begin{proof}
  Since $P \tran a Q$ and $P \tran b Q$ are coinitial and cofinal then
  they are causal equivalent. By Proposition~\ref{prop:count ceqt} the
  counting of actions  
  should be the same, hence $a=b$.
\end{proof}
UT was shown in the forward-only setting of occurrence TSIs in~\cite[Corollary~4.4]{SNW96};
it was taken as an axiom in~\cite{PU07a}. 
\begin{example}[PL alone does not imply WF or CC]\label{ex:PL not CC}
Consider the LTSI with states $P_i $ for $i = 0,1,\ldots$ and
transitions $t_i:P_{i+1} \tran a P_i$, $u_i:P_{i+1} \tran b P_i$ with $a \neq b$ and $\rev{t_i} \ind \rev{u_i}$.
BTI and SP hold.
Hence PL holds by Proposition~\ref{prop:PL}.
However clearly WF fails.
Also $t_i $ and $u_i$ are coinitial and cofinal,
and $a \neq b$, so that UT fails, and hence CC fails using Corollary~\ref{cor:ut}.
Note that the $ab$ diamonds here have the same side states so are degenerate (cf.~Lemma~\ref{lem:non-degenerate}).\finex
\end{example}
We have seen that SP is assumed when defining causal equivalence $\ceqt$.
Assuming SP, we give a diagram (Figure~\ref{fig:basic2}) to show implications between the remaining two axioms presented so far (BTI, WF) and the two main properties introduced so far (PL, CC).
We remark that the implications shown are strict (reverse implication does not hold).
We provide below counterexamples showing strictness of implications:
\begin{example}[SP, WF and CC do not imply PL]\label{ex:notPL}
  Consider the LTSI with states $P,Q,R$ and transitions $t:P\tran a R$, $u:Q \tran b R$,
  with an empty independence relation.
Then clearly BTI and PL fail.
However SP, WF and CC (and therefore UT) hold.

For CC, note that 
we can use cancellation to reduce each path to a unique shortest normal form
with respect to $\ceqt$.
There are various cases to check, depending on the initial and final states of the path, both ranging over $P,Q,R$.
Let $r:R \ptran\rho R$ be any path from $R$ to $R$.
If $r$ is non-empty, it must be of the form either $r = \rev t t r'$ or $r = \rev u u r''$.
We can use cancellation to get either $r \ceqt r'$ or $r \ceqt r''$.
Iterating the argument we see that $r \ceqt \es$. 
Now let $r:P \ptran\rho R$ be any path from $P$ to $R$.
Then $r = tr'$ where $r'$ is a path from $R$ to $R$.
Hence $r \ceqt t$.
Now let $r:P \ptran\rho P$ be any path from $P$ to $P$.
Then $r = tr'\rev t$ where $r'$ is a path from $R$ to $R$.
Hence $r \ceqt t\rev t \ceqt \es$.
Next let $r:P \ptran\rho Q$ be any path from $P$ to $Q$.
Then $r = tr'\rev u$ where $r'$ is a path from $R$ to $R$.
Hence $r \ceqt t\rev u$.
The remaining cases are similar.\finex
\end{example}
\begin{example}[SP, WF, PL and CC do not imply BTI]\label{ex:notBTI}
Consider the LTSI with states $P,Q,R,S$ and transitions $t:P\tran a Q$, $u:P \tran b R$,
$t':R \tran a S$ and $u':Q \tran b S$, with $t \ind u$.
Then BTI fails for $\rev{t'}$ and $\rev{u'}$.
However SP, WF and PL 
hold, and therefore CC also holds.

We show PL. 
As in the proof of Proposition~\ref{prop:PL}, for a path $r$ let
$d(r)$ be the number of pairs of forward transitions $(t,u)$
such that $t$ occurs to the left of $\rev u$ in $r$.
Then $r$ is parabolic iff $d(r) = 0$.

Suppose $d(r) > 0$.
We show that there is $s \ceqt r$ with $\len {s} \leq \len r$ and $d(s) < d(r)$.
Since $d(r) > 0$, we have $r = s_1 t'' \rev{u''} s_2$.
If $t'' = u''$, then we can use cancellation as in the proof of Proposition~\ref{prop:PL}. 
So suppose $t'' \neq u''$.
Since the target of $t''$ must be the same as the source of $u''$,
the only possibilities are
$t'' = t'$, $u'' = u'$ or dually $t'' = u'$, $u'' = t'$.
We consider $t'' = t'$, $u'' = u'$;
the other case is similar.
So $r = s_1 t' \rev{u'} s_2$.
Since $t \ind u$ we have $tu' \ceqt ut'$.
Hence $\rev u t u' \rev{u'} \ceqt \rev u u t' \rev{u'}$,
and so $\rev u t \ceqt t' \rev{u'}$.
So $r \ceqt s = s_1 \rev u t s_2$ and $d(s) = d(r)-1$, $\len s = \len r$.\finex
\end{example}
\begin{example}[SP and WF do not imply CC (or PL)]\label{ex:WFnotCC}
Consider the LTSI of Example~\ref{ex:notBTI},
but without $t \ind u$.
Clearly SP and WF hold.
However CC fails, since there are paths $tu'$ and $ut'$ from $P$ to $S$,
but $tu' \not\ceqt ut'$.
To see this, imagine that the four transitions of the diamond correspond to
rotations around the centre of the diamond
(see Figure~\ref{fig:WFnotCC}).
\begin{figure}[!t]
\psfrag{a}{$a$}
\psfrag{b}{$b$}
\psfrag{P}{$P$}
\psfrag{Q}{$Q$}
\psfrag{R}{$R$}
\psfrag{S}{$S$}
\begin{center}
\epsfig{file=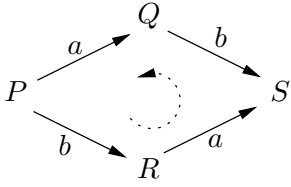, height=2.50cm}
\caption{Rotations within a diamond (Example~\ref{ex:WFnotCC}).
}\label{fig:WFnotCC}
\end{center}
\end{figure}
Measuring anti-clockwise rotation in radians
we see that $t$ and $u'$ each give a rotation of $-\pi/2$,
while $u$ and $t'$ each yield $+\pi/2$.
Let us define the rotation of a path to be the sum of the rotations of its transitions.
Path $tu'$ has rotation $-\pi$ while $ut'$ has $+\pi$.
Since there are no independent transitions,
the only operation of causal equivalence we can perform is to use
$t\rev t \ceqt \es$.
This clearly preserves the rotation of a path.
Hence $tu' \not\ceqt ut'$ as required.

PL does not hold either, otherwise CC would follow from Proposition~\ref{prop:PL WF CC}.\finex
\end{example}
\begin{example}[SP, BTI and CC do not imply WF]\label{ex:notWF}
Consider the LTSI with states $P_i $ for $i = 0,1,\ldots$ and
transitions $t_i:P_{i+1} \tran a P_i$.
Clearly WF does not hold.
However SP, BTI (and hence PL) hold; also CC (and hence UT) hold, noting that any
path is causally equivalent to a path which is entirely forward or entirely reverse.\finex
\end{example}

\section{Events}\label{sec:events}
In order to define and study causal safety and liveness (Section~\ref{sec:CSCL}),
we first need the concept of event (once further axioms are introduced, we will be able to simplify the definition below, see Definition~\ref{def:sqeqt simp}).


\begin{definition}[Event, general definition]\label{def:sqeqt}
Consider an LTSI.  
Let $\sqeqt$ be the smallest equivalence relation satisfying:
if $t:P \tran \alpha Q$, $u:P \tran \beta R$,
$u':Q \tran \beta S$, $t':R \tran \alpha S$,
and $t \ind u$, $\rev u \ind t'$, $\rev{t'} \ind \rev{u'}$, $u' \ind \rev t$,
and
\begin{itemize}
\item
$Q \neq R$ if $\alpha$ and $\beta$ are both forwards or both backwards;
\item
$P \neq S$ otherwise;
\end{itemize}
then $t \sqeqt t'$.
The equivalence classes of transitions, written $[t]$ or $[P,\alpha,Q]$, are the \emph{events}.
We say that an event is \emph{forward} if it is the equivalence class of a forward transition;
similarly for \emph{reverse} events.
Given an event $e = [t]$ we let $\rev e = [\rev t]$.
Also, we let $\und{e}=e$ if $e$ is forward, $\und{e}=\rev e$ if $e$ is backward. 
\end{definition}
Intuitively, events are the equivalence classes generated by equating transitions on the opposite sides of commuting squares.
Events are introduced as a derived notion in an LTS with independence in~\cite{SNW96},
in the context of forward-only computation.
We have changed their definition by using coinitial independence at all corners
of the diamond,
yielding rotational symmetry.
This reflects our view that forward and backward transitions have equal status. 

The labelling function $\lab$ can be extended to events
since the label does not depend on the choice of the representative inside the equivalence class.


\subsection{Pre-reversible LTSIs}

Our definition of event can be simplified if the LTSI, and independence in
particular, are well-behaved. Thus, we now add a further axiom related
to independence. This leads us to pre-reversible LTSIs.

\begin{definition}\label{def:PCI}
  {\bf Propagation of coinitial independence (PCI)}\footnote{PCI was called CPI (coinitial propagation of independence) in~\cite{LanesePU20}; we changed the terminology following a suggestion from Marco Bernardo to better match the intuition.}:
  if $t:P
    \tran \alpha Q$, $u:P \tran \beta R$, $u': Q \tran \beta S$ and
    $t':R \tran \alpha S$ with $t \ind u$, then $u' \ind \rev t$.
\end{definition}

PCI
states that independence is a property of commuting diamonds more than
of their specific pairs of edges. Indeed,
it allows
independence to propagate around a commuting diamond.

\begin{definition}[Pre-reversible LTSI]\label{def:prerev}
If an LTSI satisfies axioms SP, BTI, WF and PCI,
we say that it is \emph{pre-reversible}.
\end{definition}
The name `pre-reversible' indicates that we expect to require further axioms,
but the present four are enough to ensure that LTSIs are well-behaved,
with events compatible with causal equivalence (Lemma~\ref{lemma:cccount}). Pre-reversible axioms are separated from further
axioms by a dashed line in Table~\ref{t:list}.

A first consequence of PCI is that coinitial transitions with
labels $\alpha$ and $\rev \alpha$ are not independent.
\begin{lemma}\label{lemma:revnotind}
Suppose that an LTSI satisfies PCI. 
If $t: P \tran\alpha Q$ and $u: P \tran{\rev\alpha} R$ are coinitial transitions, 
then $t \notind u$.
\end{lemma}
\begin{proof}
Suppose that $t: P \tran\alpha Q$ and $u: P \tran{\rev\alpha} R$
are independent.
Consider the degenerate diamond with two copies of $P$ and transitions $t,u,\rev t, \rev u$.
By applying PCI we deduce $\rev t \ind \rev t$, which contradicts irreflexivity of $\ind$.
\end{proof}
Additionally, we cannot have two different coinitial backward transitions with the same label.
\begin{definition}\label{def:BD}
{\bf Backward label determinism (BLD):}:
if
$t: P \rtran a Q$ and $u: P \rtran a R$ are coinitial backward transitions
with the same label then $t = u$.
\end{definition}
\begin{proposition}\label{prop:BD}
Suppose that an LTSI satisfies SP, BTI and PCI. Then it satisfies BLD.
\end{proposition}
\begin{proof}
Suppose $t: P \rtran a Q$ and $u: P \rtran a R$.
Then if $t \neq u$ we have $t \ind u$ by BTI.
We can complete a diamond with $Q \rtran a S$, $R \rtran a S$ by SP.
But then $Q \ftran a P$ and $Q \rtran a S$ are independent by PCI.
This is a contradiction of Lemma~\ref{lemma:revnotind}.
\end{proof}
A consequence of Lemma~\ref{lemma:revnotind} is that an LTSI
satisfying BTI and PCI cannot include a diamond
$P \tran \alpha Q \tran \alpha S$, $P \tran \alpha R \tran \alpha S$
where all four transitions have the same label.
This can be seen as ruling out \emph{autoconcurrency}~\cite{Bed91}.

The following non-degeneracy property was shown for occurrence transition systems with independence in~\cite[page~312]{SNW96}, which considers forward transitions only.
We have to cope with backward as well as forward transitions.
%
\begin{lemma}\label{lem:non-degenerate}
Suppose that an LTSI is pre-reversible.
If we have a diamond
$t:P \tran \alpha Q$, $u:P \tran \beta R$ with $t \ind u$
together with cofinal transitions $u': Q \tran \beta S$ and $t': R \tran\alpha S$,
then the diamond is \emph{non-degenerate},
meaning that $P,Q,R,S$ are distinct states.
\end{lemma}
%
\begin{proof}
We note that CC holds; hence UT holds thanks to Corollary~\ref{cor:ut}.
By WF we see that $P \neq Q \neq S \neq R \neq P$.
It remains to show $Q \neq R$ and $P \neq S$.

Suppose $Q = R$.
By $t \ind u$ we know $t \neq u$.  So $\alpha \neq \beta$.
But if $\alpha$ and $\beta$ are both forward or both backward this is impossible by UT.
If one is forward and the other is backward then this is impossible by WF.
Hence $Q \neq R$.

Suppose $P = S$.
If $\alpha$ and $\beta$ are both forward or both backward this is impossible by WF.
If one is forward and the other is backward then by UT this implies that
$\alpha = \rev\beta$.
Then $t \notind u$ by Lemma~\ref{lemma:revnotind},
which is a contradiction.
Hence $P \neq S$.
\end{proof}

If an LTSI is pre-reversible then by
Lemma~\ref{lem:non-degenerate} and the use of PCI
we can simplify the statement of Definition~\ref{def:sqeqt}
to:

\begin{definition}[Event, simplified definition]\label{def:sqeqt simp}
  Consider a pre-reversible LTSI.
Let $\sqeqt$ be the smallest equivalence relation satisfying:
if $t:P \tran \alpha Q$, $u:P \tran \beta R$,
$u':Q \tran \beta S$, $t':R \tran \alpha S$,
and $t \ind u$,
then $t \sqeqt t'$.
\end{definition}

We are now able to show independence of diamonds (ID), which can be seen as
dual of SP.

\begin{definition}\label{def:ID}
%
%
{\bf Independence of diamonds (ID)}: if we have a diamond
$t:P \tran \alpha Q$, $u:P \tran \beta R$,
$u': Q \tran \beta S$ and $t':R \tran \alpha S$,
with 
\begin{itemize}
\item
$Q \neq R$ if $\alpha$ and $\beta$ are both forwards or both backwards;
\item
$P \neq S$ otherwise;
\end{itemize}
then $t \ind u$.
\end{definition}
\begin{proposition}\label{prop:ID}
    If an LTSI satisfies BTI and PCI then it satisfies ID.
\end{proposition}
\begin{proof}
Suppose we have a diamond
$t:P \tran \alpha Q$, $u:P \tran \beta R$,
$u': Q \tran \beta S$ and $t':R \tran \alpha S$,
with 
\begin{itemize}
\item
$Q \neq R$ if $\alpha$ and $\beta$ are both forwards or both backwards;
\item
$P \neq S$ otherwise.
\end{itemize}
We must show $t \ind u$.
There are various cases, depending on whether $\alpha$ and $\beta$ are forwards or backwards.
If they are both forwards, then $Q \neq R$.  Hence $\rev{t'} \neq \rev{u'}$
and by BTI we have $\rev{t'} \ind \rev{u'}$.
By PCI, $u' \ind \rev t$ and again by PCI $t \ind u$ as required. 
Other cases are similar.
\end{proof}
In the proof of the above proposition it must be the case that
$\und\alpha \neq \und\beta$, or else we get a contradiction using
Lemma~\ref{lemma:revnotind}.
  

\subsection{Counting occurrences of events}
We now consider the interaction between events and causal equivalence.
We need some notation first.

\begin{definition}\label{def:count events}
  Let $r$ be a path and $e$ be an event of the same LTSI.
Let $\cte(r,e)$ be the number of occurrences of transitions $t$ in $r$
such that $t \in e$, minus the number of occurrences of transitions $t$ in $r$ such that $t \in \rev e$.
We define $\cte(r,e)$ by induction on the length of $r$ as follows:
\begin{equation*}
\begin{split}
\cte(\es,e) & = 0 \\
\cte(tr,e) & =
\begin{cases}
\cte(r,e)+1 & \text{if } [t] = e \\
\cte(r,e)-1 & \text{if } [t] = \rev e \\
\cte(r,e) & \text{otherwise}
\end{cases}
\end{split}
\end{equation*}
\end{definition}

We now show that $\cte(r,e)$ is invariant under causal equivalence.

\begin{lemma}\label{lemma:cccount}
  Assume an LTSI is pre-reversible.
  Let $r \ceqt s$. Then for each event $e$ we have that $\cte(r,e) = \cte(s,e)$.
\end{lemma}
\begin{proof}
  We prove the thesis for $r$ and $s$ being derived by a single
  application of the axioms; the thesis will follow since equality is
  an equivalence relation.

  If $r=r_1tu'r_2$ and $s=r_1ut'r_2$ then we have by
  definition of causal equivalence (Definition~\ref{def:ceqt})
  that
  $t \ind u$.
Hence,
  $[t]=[t']$ and $[u]=[u']$ using Definition~\ref{def:sqeqt simp}. The thesis follows.

  If $r=r_1t\rev tr_2$ and $s=r_1r_2$ (the other case is analogous)
  then the contribution of $t$ and $\rev t$ to $\cte(r,[t])$ (as well
  as to $\cte(r,e)$ for $t \notin e$) is $0$; hence the thesis
  follows.
\end{proof}
Lemma~\ref{lemma:cccount} generalises what was shown for the forward-only setting
in~\cite[Corollary~4.3]{SNW96}.
%
\begin{proposition}\label{prop:regeqzero}
If an LTSI is pre-reversible,
then for any rooted path $r$ and any forward event $e$ we have $\cte(r,e) \geq 0$.
\end{proposition}
\begin{proof}
Let $r$ be a rooted path.
Using PL (Proposition~\ref{prop:PL}), we obtain a coinitial and cofinal forward-only path $s$ such that $s \ceqt r$.  Let $e$ be any forward event.  Clearly $\cte(s,e) \geq 0$.
Hence $\cte(r,e) \geq 0$ by Lemma~\ref{lemma:cccount}.
\end{proof}
%
We can lift independence from transitions to events.

\begin{definition}[Coinitially independent events]\label{def:coind events}
Let events $e,e'$ be \emph{coinitially independent},
written $e \coind e'$, iff there are coinitial transitions $t,t'$ such that
$[t] = e$, $[t'] = e'$ and $t \ind t'$.
\end{definition}
\begin{lemma}\label{lem:coind rev}
Assume an LTSI is pre-reversible. If $e \coind e'$ then we have also
$\rev e \coind e'$.
\end{lemma}
\begin{proof}
Suppose that $e \coind e'$.
Then there are coinitial $t, u$ such that $[t] = e$, $[u] = e'$ and $t \ind u$.
Use SP to complete a diamond with transitions $t' \sqeqt t$, $u' \sqeqt u$.
By PCI we have $\rev t \ind u'$.
Hence $\rev e \coind e'$ as required.
\end{proof}
Thus in pre-reversible LTSIs, $\coind$ is fully determined just considering forward events.
By Lemma~\ref{lem:coind rev},
if we know $e \coind e'$ then we know $\und e \coind \und{e'}$.
\begin{proposition}\label{prop:coind irref}
Assume an LTSI is pre-reversible.  Then $\coind$ is irreflexive.
\end{proposition}
\begin{proof}
Suppose for a contradiction that $e \coind e$ for some event $e$.
By Lemma~\ref{lem:coind rev},
we can assume that $e$ is forward.
Then there are coinitial transitions $t,u \in e$ such that $t \ind u$.
We can use SP to complete a square with $t' \sqeqt t$ and $u' \sqeqt u$.
This square is non-degenerate by Lemma~\ref{lem:non-degenerate}.
But now $\rev{t'}$ and $\rev{u'}$ are two distinct coinitial backward transitions with the same label, contradicting BLD (Proposition~\ref{prop:BD}).
\end{proof}
We can slightly strengthen the previous result as follows:
\begin{proposition}\label{prop:coind und}
Assume an LTSI is pre-reversible.
If $t:P \tran\alpha Q$ and $u:R \tran\beta S$ with $[t] \coind [u]$
then $\und\alpha \neq \und\beta$.
\end{proposition}
\begin{proof}
Similar to the proof of Proposition~\ref{prop:coind irref}.
\end{proof}

In pre-reversible LTSIs each event can occur at most once in a rooted path.
\begin{definition}\label{def:NRE}
{\bf No repeated events (NRE)}: for
any rooted path $r$ and any forward event $e$ we have $\cte(r,e) \leq 1$.
\end{definition}
In order to prove NRE we need the following lemmas.
\begin{lemma}[Ladder Lemma]\label{lem:ladder}
Assume an LTSI is pre-reversible.
Suppose that $t:P \tran \alpha Q$ and $t':P' \tran \alpha Q'$ with $t \sqeqt t'$.
Then there is a path $s$ from $Q$ to $Q'$ such that for all $u$ in $s$
we have $[t] \coind [u]$.
\end{lemma}
\begin{figure}[!t]
\psfrag{t}{$t$}
\psfrag{t'}{$t'$}
\psfrag{t''}{$t''$}
\psfrag{u}{$u$}
\psfrag{u'}{$u'$}
\psfrag{P}{$P$}
\psfrag{Q}{$Q$}
\psfrag{P'}{$P'$}
\psfrag{Q'}{$Q'$}
\begin{center}
\epsfig{file=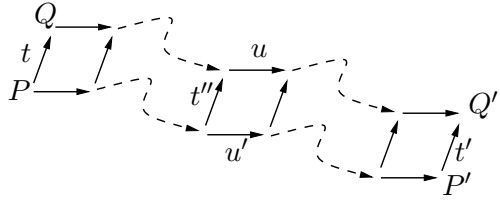, height=2.50cm}
\caption{The ladder of diamonds in the proof of Lemma~\ref{lem:ladder}.
}\label{fig:ladder}
\end{center}
\end{figure}
\begin{proof}
By the definition of $\sqeqt$ there is a ladder of diamonds (see Figure~\ref{fig:ladder})
connecting $t$ to $t'$.
This gives a path $s$ from $Q$ to $Q'$.
Take any $u$ in $s$, and consider the diamond containing $u$.
Let $u'$ be on the opposite side from $u$, so that $u' \sqeqt u$,
and let $t''$ be the rung nearest to $t$, so that $t \sqeqt t''$.
We have $t'' \ind u'$.  Hence, the result follows.
\end{proof}
\begin{lemma}\label{lem:cte zero}
Let $\mc L$ be a pre-reversible LTSI.
Suppose $t: P \tran\alpha Q$ and $t': P' \tran\alpha Q'$
with $t \sqeqt t'$, and suppose $r$ is a path from $Q$ to $Q'$.
Then $\cte(r,[t]) = 0$.
\end{lemma}
\begin{proof}
By Lemma~\ref{lem:ladder} there is a path $s$ from $Q$ to $Q'$
such that for all $u$ in $s$ we have $[t] \coind [u]$.
Let 
$\lab(u)=\beta$.
By Proposition~\ref{prop:coind und} we have  $\und\alpha \neq \und\beta$.
Hence $\cte(s,[t]) = 0$,
and by Lemma~\ref{lemma:cccount} $\cte(r,[t]) = 0$ as required.
\end{proof}

\begin{proposition}\label{prop:NRE}
If an LTSI is pre-reversible then it satisfies NRE.
\end{proposition}
\begin{proof}
Let $e$ be a forward event and $r$ be a rooted path from $I$ to $R$, and suppose for a contradiction that
$\cte(r,e) > 1$.
Using PL we can obtain a forward-only path $r'$ from $I$ to $R$ with $r \ceqt r'$.
By Lemma~\ref{lemma:cccount}, $\cte(r',e) > 1$.
Hence, $r'$ contains (at least) two transitions for $e$; let us denote them as
$t:P \tran a Q$ and $t':P' \tran a Q'$. Without loss of generality, we assume that $t$ occurs before $t'$ in~$r$. 
Let $r''$ be the portion of $r'$ from $Q$ to $P'$.
By Lemma~\ref{lem:cte zero} applied to $t,t'$ and path $r''t'$
we have $\cte(r''t',[t]) = 0$.
This is a contradiction since $r''$ is forward-only.
\end{proof}
NRE was shown in the forward-only setting of occurrence transition systems with independence in~\cite[Corollary~4.6]{SNW96}.
It was also shown in the reversible setting without independence
in~\cite[Proposition~2.10]{PU07a}.

\begin{example}\label{ex:repeated}
Consider the LTSI in Figure~\ref{fig:repeated}.
\begin{figure}[!t]
\psfrag{a}{$a$}
\psfrag{b}{$b$}
\begin{center}
\epsfig{file=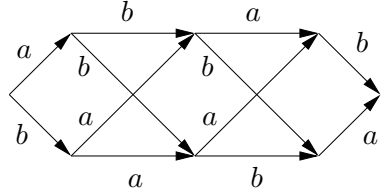, height=2.50cm}
\caption{The LTSI in Example~\ref{ex:repeated}.
}\label{fig:repeated}
\end{center}
\end{figure}
Independence holds only between coinitial transitions and is given by closing under BTI and propagating
independence around the corners of diamonds as in PCI whenever possible.
Note, however, that PCI does not hold, since we have coinitial independent $a$ and $\rev a$-transitions,
contradicting Lemma~\ref{lemma:revnotind}.
On the other hand, in addition to BTI, axioms SP and WF hold, so that CC holds.
All $a$-transitions belong to the same event,
and all $b$-transitions belong to the same event.
We have rooted paths where the same event is repeated,
contradicting NRE.
Note also that BLD fails and that $\coind$ is reflexive.\finex
\end{example}

\subsection{Polychotomy}
We now show what we call \emph{polychotomy},
which states that if forward events do not cause each other and are not in conflict,
then they must be independent.
This will help us to relate the different notions of causal safety and liveness (Section~\ref{sec:CSCL}).
We first define causality and conflict relations on forward events.
\begin{definition}[Causality relation on forward events]\label{def:ordering}
Let $e,e'$ be forward events of some LTSI. 
Let $e \leq e'$ iff
for all rooted paths $r$, if $\cte(r,e') > 0$ then $\cte(r,e) > 0$.
As usual $e < e'$ means $e \leq e'$ and $e \neq e'$.
If $e < e'$ we say that
$e$ is a \emph{cause} of $e'$.
\end{definition}
As expected, the causality relation is a partial ordering (i.e., a reflexive, transitive and antisymmetric relation).
\begin{lemma}\label{lem:po}
If an LTSI is pre-reversible then
$\leq$ is a partial ordering on events.
\end{lemma}
\begin{proof}
Reflexivity and transitivity are immediate.
For antisymmetry, suppose that $e_1 \leq e_2$ and $e_2 \leq e_1$,
where $e_1,e_2$ are forward events.
Then for all rooted $r$, $\cte(r,e_1) >0$ iff $\cte(r,e_2) >0$.
Since the LTSI is pre-reversible, by Proposition~\ref{prop:regeqzero},
for all rooted $r$, $\cte(r,e_1) \geq 0$ and $\cte(r,e_2) \geq 0$.
Let $r$ be a shortest rooted path such that $\cte(r,e_1) >0$.
We can use WF to show that $r$ must exist.
Then $\cte(r,e_2) > 0$.
Also $r = r't$, where $\cte(r',e_1) = 0$ (otherwise $r$ would not be a shortest path) and so $\cte(r',e_2) = 0$.
We see that both $[t] = e_1$ and $[t] = e_2$,
showing that $e_1 = e_2$ as required.
\end{proof}
In~\cite{vGV97,PU07a}, orderings on forward events have been defined using forward-only rooted paths;
in fact, the definitions coincide for pre-reversible LTSIs.
\begin{definition}[\cite{vGV97,PU07a}]\label{def:ordering fwd}
Let $e,e'$ be forward events of some LTSI. 
Let $e \leqf e'$ iff
for all forward-only rooted paths $r$,
if  $\cte(r,e') >0$
then $\cte(r,e) >0$.
\end{definition}
\begin{lemma}\label{lem:ordering}
For any LTSI, and any forward events $e,e'$,
$e \leq e'$ implies $e \leqf e'$.
If an LTSI is pre-reversible then
$e \leqf e'$ implies $e \leq e'$.
\end{lemma}
\begin{proof}
Straightforward using PL and Lemma~\ref{lemma:cccount}.
\end{proof}
\begin{definition}\label{def:conflict}
Two forward events $e,e'$ are in \emph{conflict}, written $e \cf e'$,
iff there is no rooted path $r$ such that $\cte(r,e) >0$ and $\cte(r,e') > 0$.
\end{definition}
Much as for orderings, conflict on events has been defined previously
using forward-only rooted paths~\cite{vGV97,PU07a};
in fact, the definitions coincide for pre-reversible LTSIs.
We omit the details.

We can now introduce the main result of this section.
\begin{definition}[Polychotomy]\label{def:poly}
Let $\mc L$ be a pre-reversible LTSI.
We say that $\mc L$ satisfies \emph{polychotomy} if whenever
$e,e'$ are \emph{forward} events, then exactly one of the following holds: 
\begin{enumerate}
\item $e = e'$;\quad
\item $e < e'$;\quad
\item $e' < e$;\quad
\item $e \cf e'$; or\quad 
\item $e \coind e'$. 
\end{enumerate}
\end{definition}
%
\begin{proposition}[Polychotomy]\label{prop:poly}
Assume an LTSI is pre-reversible.
Then polychotomy holds.
\end{proposition}
\begin{proof}
Consider two forward events $e$ and $e'$ which may or may not be equal.

We first check mutual exclusivity.
Suppose $e = e'$.
Then $e < e$ is impossible by definition of $<$.
Also $e$ cannot be in conflict with itself (we can use WF to show that there is at least one rooted path).
Finally, $e \coind e$ is impossible by Proposition~\ref{prop:coind irref}.
From now on we assume $e \neq e'$.

Next suppose $e < e'$.
We can rule out $e' < e$ using 
Lemma~\ref{lem:po}.

Using Lemma~\ref{lem:ordering}, we know that $e \ltf e'$, hence there must be some forward-only rooted path with $e$ followed by $e'$ (WF ensures at least one rooted path exists),
and so $e$ and $e'$ are not in conflict.
Finally $e \coind e'$ implies that there are two coinitial transitions $t \in e$, $t' \in e'$
which are independent.
Using SP to complete the square we see that $e < e'$ is impossible by NRE,
which holds by Proposition~\ref{prop:NRE}.

Similarly we see that $e' < e$ implies that $e$ and $e'$ are not in conflict and not independent.

Next suppose that $e \cf e'$.
If $e \coind e'$
then there are two coinitial transitions $t \in e$, $t' \in e'$
which are independent.
Using SP to complete the square and WF we see that we have a forward-only rooted path
containing occurrences of both $e$ and $e'$ contradicting them being in conflict.

Suppose that none of (1)-(4) hold.
We must show (5).
Since $e,e'$ do not conflict, there is a
rooted path $r$ starting at some irreversible $I$ such that $\cte(r,e) > 0$ and $\cte(r,e') > 0$. If more than one such path exists, choose one of minimal length.
 W.l.o.g.~suppose that $r$ finishes with $t' \in e'$ at $P$.
Since not $e < e'$, using Lemma~\ref{lem:ordering} also $e \ltf e'$ does not hold; hence there is another forward-only path~$r'$ from some irreversible $I'$
finishing with $t'' \in e'$ at $Q$
such that $\cte(r',e) = 0$.
By Lemma~\ref{lem:ladder} there is a path $s$ from $Q$ to $P$
such that $e' \coind [u]$ for every $u$ in $s$.
Using Proposition~\ref{prop:unique irrev} we deduce that $I' = I$.
By CC $r \ceqt r's$ and so by Lemma~\ref{lemma:cccount} $\cte(s,e) > 0$ and $s$ must contain $u \in e$,
yielding $e \coind e'$ as required.
\end{proof}

\section{Causal Safety and Causal Liveness}\label{sec:CSCL}
\begin{figure}[!t]
\psfrag{IRE+IEC}{IRE+IEC}
\psfrag{IRE+RPI}{IRE+RPI}
\psfrag{CIRE+IEC}{CIRE+IEC}
\psfrag{IEC}{IEC}
\psfrag{IRE}{IRE}
\psfrag{CIRE}{CIRE}
\psfrag{IC}{IC}
\psfrag{IC+CIRE}{IC+CIRE}
\psfrag{RPI}{RPI}
\psfrag{CSi}{CS$\indt$}
\psfrag{CSi+RPI}{CS$\indt$+RPI}
\psfrag{CLi}{CL$\indt$}
\psfrag{CLci}{CL$\ci$}
\psfrag{IC+CLci}{IC+CL$\ci$}
\psfrag{IEC+CLci}{IEC+CL$\ci$}
\psfrag{LG}{LG}
\psfrag{LG+IEC}{LG+IEC}
\psfrag{CLG}{CLG}
\psfrag{ECh}{ECh}
\psfrag{EIT}{EIT}
\psfrag{IEC+CLci}{IEC+CL$\ci$}
\psfrag{LGIEC}{Ex.~\ref{ex:LG+IEC}} %
\psfrag{eLG}{Ex.~\ref{ex:LG}} 
\psfrag{eIRE+IEC}{Ex.~\ref{ex:IRE+IEC}} 
\psfrag{CSi+RPICLi}{Ex.~\ref{ex:CSi+RPI CLi}} 
\psfrag{IRE1}{Ex.~\ref{ex:IRE1}} 
\psfrag{IRE2}{Ex.~\ref{ex:IRE2}} 
\psfrag{CLGCSi}{Ex.~\ref{ex:CLG CSi}} 
\psfrag{prerevnotCL}{Ex.~\ref{ex:prerev not CL}} %
\psfrag{halfcubemod}{Ex.~\ref{ex:halfcube mod}} %
\psfrag{eCLG}{Ex.~\ref{ex:prerev not CSi} } 
\psfrag{eIC+CIRE}{Ex.~\ref{ex:IC+CIRE}} 
\psfrag{halfcube}{Ex.~\ref{ex:halfcube}} %
\psfrag{ICCLi}{Ex.~\ref{ex:IC CLi}} 
\psfrag{eIC}{Ex.~\ref{ex:IC}} 

\begin{center}
\epsfig{file=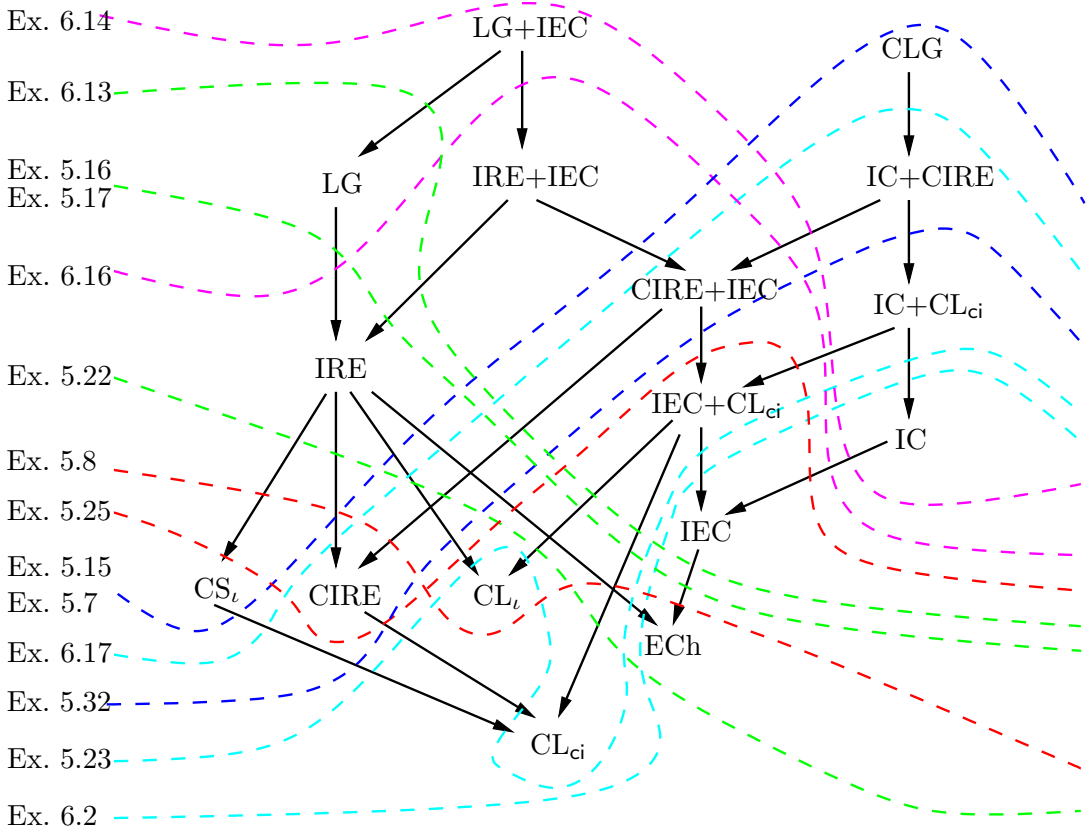, height=11.0cm}
\caption{Implications between properties all assuming pre-reversible.
Note that CS$\ci$ holds for pre-reversible LTSIs (Theorem~\ref{thm:CS coind}).
All implications are strict, and there are no further implications between the sixteen properties shown,
in view of the examples cited.
}\label{fig:diagprerev1}
\end{center}
\end{figure}

In the literature, causal consistent reversibility is frequently
informally described by saying that ``a transition can be undone if
and only if each of its consequences, if any, has been undone''
%
(see, e.g., \cite{LaneseNPV18}).
In this section we study this
property, where the two implications will be referred to as \emph{causal
  safety} and \emph{causal liveness}. We provide three different formalisations 
of
such properties, based on
independence of transitions (Section~\ref{sub:indtra}),
independence of events (Section~\ref{sub:indev}), and
ordering of events (Section~\ref{sub:ord}),
and study their relationships.
In Figure~\ref{fig:diagprerev1} we show the relationships
between the various axioms and properties we shall study
in this section and Section~\ref{sec:coinitial}.
\subsection{CS and CL via independence of transitions}\label{sub:indtra}





We first define causal safety and liveness using the independence relation.
\begin{definition}\label{def:safe live}
Let $\mc L$ be an LTSI. 
\begin{enumerate}
\item
We say that $\mc L$ is \emph{causally safe (CS$\indt$)} if whenever
$t_0:P \tran a Q$, $r:Q \ptran \rho R$, $\cte(r,[t_0]) = 0$ and $t_0\op:S \tran a R$
with $t_0 \sqeqt t_0\op$,
then $\rev{t_0} \ind t$ for all $t$ in $r$ such that $\cte(r,[t]) > 0$.
%
%
\item
We say that $\mc L$ is \emph{causally live (CL$\indt$)} if whenever
$t_0:P \tran a Q$, $r:Q \ptran \rho R$ and $\cte(r,[t_0]) = 0$ and
$\rev{t_0} \ind t$, for all $t$ in $r$ such that $\cte(r,[t]) > 0$,
then we have
$t_0\op:S \tran a R$ with $t_0 \sqeqt t_0\op$.
\end{enumerate}
\end{definition}
Properties CS$\indt$ and CL$\indt$ both consider a (forward) transition $t_0:P \tran a Q$  
followed by a path $r$ where the number of occurrences in $r$
of transitions that belong to the same event as $t_0$ is zero.
CS$\indt$ states that if after path $r$ a transition $t_0\op$ can be undone, where
$t_0$ and $t_0\op$ belong to the same event, then the reverse of $t_0$ 
is independent of all transitions $t$ where the number of occurrences in $r$ of 
the event of $t$ is positive.
Dually, CL$\indt$ requires that if the reverse of $t_0$ is independent of all transitions 
whose events have a positive number of occurrences in $r$, then it can be undone.

%

\begin{remark}\label{rem:equal0notneeded}
In the definition of CS$\indt$ the condition that $\cte(r,[t_0]) = 0$
can be deduced from the other conditions using
Lemma~\ref{lem:cte zero},
provided that the LTSI is pre-reversible.
\end{remark}

We use the reverse of $t_0$ when considering independence from $t$
because our axioms BTI, SP and PCI focus on \emph{coinitial} independence
rather than independence of consecutive transitions in a trace.
Take the simplest case where $r$ is a single transition $t:Q \tran b R$.
First assume $\rev{t_0} \ind t$;
note that this is coinitial independence.
We can use SP and PCI to get $t_0\op:S \tran a R$ with $t_0 \sqeqt t_0\op$,
which is an example of causal liveness.
Conversely, if we assume $t_0\op:S \tran a R$ with $t_0 \sqeqt t_0\op$,
we can use BTI, SP, BLD and PCI to get a diamond with $\rev{t_0} \ind t$,
which is an example of causal safety.

Note that in the discussion above to prove causal safety we need to consider also the case $r=t \rev t t$. Since $[\rev t]$ has a negative number of occurrences, we only need to show that $\rev{t_0} \ind t$, which can be proved as above. However, if we replaced the condition $\cte(r,[t]) > 0$ with $\cte(r,[t]) \neq 0$, we would also need to show  $\rev{t_0} \ind \rev t$, which does not follow from the axioms above. Intuitively, requiring $\rev{t_0} \ind \rev t$ would make little sense, since all the occurrences of $\rev t$ could be simplified with corresponding occurrences of $t$. This is why we decided to require $\cte(r,[t]) > 0$.


We have seen in the last two paragraphs that existing axioms are sufficient to show
CS$\indt$ and CL$\indt$ in the case where trace $r$ consists of
a single transition. However, existing axioms are not enough
for general $r$, as we will show in Examples~\ref{ex:prerev not CSi} and~\ref{ex:prerev not CL}.
Thus, we introduce
the following
axiom, which states that independence does not depend on the choice
of the representative inside an event.
\begin{definition}\label{def:IRE}
  {\bf Independence respects events (IRE)}:
Whenever $t \sqeqt t' \ind u$ we have $t \ind u$.
\end{definition}
IRE is one of the conditions in the definition of transition systems with
independence~\cite[Definition~3.7]{SNW96}.

IRE allows us to relate coinitial independence on events and independence on transitions.
\begin{lemma}\label{lem:coind IRE}
Assume an LTSI satisfies IRE.
If $[t] \coind [u]$ then $t \ind u$.
\end{lemma}
\begin{proof}
Immediate.
\end{proof}

Together with the axioms for pre-reversibility,
IRE is enough to show both CS$\indt$ and CL$\indt$.

\begin{theorem}\label{thm:CS}
Let a pre-reversible LTSI satisfy IRE.
Then it satisfies CS$\indt$.
\end{theorem}
\begin{proof}
Suppose $t_0:P \tran a Q$, $r:Q \ptran \rho R$ and $t_0\op:S \tran a R$
with $t_0 \sqeqt t_0\op$.
By Lemma~\ref{lem:ladder} there is a path $s$ from $Q$ to $R$
such that for all $u$ in $s$ we have $[t_0] \coind [u]$.
We deduce by Lemmas~\ref{lem:coind rev} and~\ref{lem:coind IRE}
that for all $u$ in $s$ we have 
$\rev{t_0} \ind u$. 
By CC, $r \ceqt s$.

Take $t$ in $r$ such that $\cte(r,[t]) > 0$.
Then $\cte(s,[t]) > 0$, thanks to Lemma~\ref{lemma:cccount}.
But then there is $u$ in $s$ such that $u \sqeqt t$.
We have $\rev{t_0} \ind u$ and so $\rev{t_0} \ind t$, using IRE, as desired.
\end{proof}

%

\begin{theorem}\label{thm:CL}
  Let a pre-reversible LTSI satisfy IRE.
  Then it satisfies CL$\indt$.  
\end{theorem}
%
%
%
  %
%
\begin{proof}
  Suppose $t_0:P \tran a Q$, $r:Q \ptran \rho R$ and $\cte(r,[t_0]) =
  0$ and $t_0 \ind t$, for all $t$ in $r$ such that $\cte(r,[t]) > 0$.  We
  have to show that there is $t_0\op:S \tran a R$ with $t_0 \sqeqt
  t_0\op$.

  Thanks to PL, there is $T$ such that $b: P
  \ptran{\rho_b} T$ and $f: T \ptran{\rho_f} R$, with $b$ backward and
  $f$ forward.
  By CC, $t_0r \ceqt bf$.
  Since $\cte(r,[t_0]) = 0$, thanks to
  Lemma~\ref{lemma:cccount} we have $\cte(bf,[t_0])=1$. As a consequence,
  there is a transition $t'_0:P' \tran a Q' \in [t_0]$ in
  $f$.
This $t'_0$ is in fact the unique transition in $[t_0]$ belonging to $f$ by Proposition~\ref{prop:NRE}.
  Let $f'$ be the portion of $f$ from $Q'$ to $R$.
  If we can show that $\rev{t'_0} \ind t''$ for each transition $t''$ in $f'$,
  then the thesis will follow by commuting
  $t'_0$ with all such transitions using SP and IRE.

  By Lemma~\ref{lem:ladder} there is a path $s$ from $Q$ to $Q'$ such that $[t_0] \coind [u]$
  for all $u$ in $s$.
  By CC, $r \ceqt sf'$.
  Take any $t''$ in $f'$.
  By Lemma~\ref{lemma:cccount}, $\cte(r,[t'']) = \cte(s,[t'']) + \cte(f',[t''])$.
  If $\cte(s,[t'']) < 0$ then there is $u$ in $s$ such that $u \sqeqt \rev{t''}$.
Now $[t_0] \coind [u] = [\rev{t''}]$.
Therefore $[\rev{t_0}] \coind [t'']$ by Lemma~\ref{lem:coind rev}, and $\rev{t_0} \ind t''$ by Lemma~\ref{lem:coind IRE}.
  Suppose instead $\cte(s,[t'']) \geq 0$.
  Since $\cte(f',[t'']) > 0$, we have $\cte(r,[t'']) > 0$. 
  So there is $u$ in $r$ such that $u \sqeqt t''$,
  and by hypothesis $\rev{t_0} \ind u$, so that $\rev{t'_0} \ind t''$ using IRE.
\end{proof}

We now give examples of LTSIs which are pre-reversible and where CS$\indt$ and CL$\indt$ fail.
\begin{example}\label{ex:prerev not CSi}
Consider the LTSI shown in Figure~\ref{fig:notIRE} including the dashed transitions.
We add coinitial independence as generated by BTI and PCI.
BTI gives $(Q',\rev b,Q) \ind (Q',\rev a,P')$ and  $(R,\rev c,Q') \ind (R,\rev a,S)$.
Assuming $t_0:P \tran a Q$ and $t_0\op:S \tran a R$, PCI gives three additional independence pairs for each of
 the two diamonds:  $(Q, b,Q') \ind \rev{t_0}$, $t_0 \ind (P,b,P')$ and $(P', \rev{b}, P) \ind (P', a, Q')$  for the diamond with the source $P$, and  $(Q', c,R) \ind (Q', \rev{a}, P')$, $(P',a,Q') \ind (P',c,S)$ and $(S,\rev{c}, P') \ind t_0\op $ for the other diamond.
The LTSI is pre-reversible.
However CS$\indt$ fails.
Transition $t_0$ is followed by a path $Q \ptran {bc} R$ and the transition $t_0\op$
satisfies $t_0 \sqeqt t_0\op$.
If CS$\indt$ held we could deduce that $\rev{t_0} \ind (Q',c,R)$,
which is not the case.
Similarly, we see that IRE fails, since
$\rev{t_0} \sqeqt (Q',\rev a ,P') \ind (Q',c,R)$ but not $\rev{t_0} \ind (Q',c,R)$.
Note, however, that CL$\indt$ holds, since only transitions inside the same diamond are independent, and transitions on one side of the diamond are undone by the corresponding transition on the opposite side.
\finex
\begin{figure}[!t]
\psfrag{a}{$a$}
\psfrag{b}{$b$}
\psfrag{c}{$c$}
\psfrag{P}{$P$}
\psfrag{P'}{$P'$}
\psfrag{Q}{$Q$}
\psfrag{Q'}{$Q'$}
\psfrag{R}{$R$}
\psfrag{S}{$S$}
\begin{center}
\epsfig{file=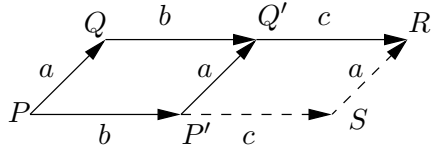, height=2cm}
\caption{The LTSIs in Examples~\ref{ex:prerev not CSi} and~\ref{ex:prerev not CL}.
}\label{fig:notIRE}
\end{center}
\end{figure}
\end{example}

\begin{example}\label{ex:prerev not CL}
Consider the LTSI shown in Figure~\ref{fig:notIRE} excluding the dashed transitions.
We add coinitial independence as given by BTI and PCI, similarly to the previous example.
We also add $(Q,\rev a ,P) \ind (Q',c,R)$.
The LTSI is pre-reversible.
However CL$\indt$ fails.
We have $t_0:P \tran a Q$, $Q \ptran {bc} R$ and $\rev{t_0} \ind (Q,b,Q')$, $\rev{t_0} \ind (Q',c,R)$.
Clearly CL$\indt$ fails, since we cannot reverse the $a$-transition at $R$. 
IRE fails since $(Q',\rev a ,P') \sqeqt \rev{t_0} \ind (Q',c,R)$
but not $(Q',\rev a ,P') \ind (Q',c,R)$.
Note, however, that CS$\indt$ holds since the only way to undo transitions is with transitions on the opposite side of the same diamond, and the path connecting them is another transition of the same diamond. Hence, the condition on independence holds, thanks to BTI and PCI.
\finex
\end{example}
Examples~\ref{ex:prerev not CSi} and~\ref{ex:prerev not CL} show that the stipulation of IRE cannot be omitted in the statements of Theorems~\ref{thm:CS} and~\ref{thm:CL}, respectively.
These examples also show that we cannot deduce CS$\indt$ or CL$\indt$ from CC, nor one from the other.
\begin{example}[CS$\indt$ and CL$\indt$ do not imply CC]\label{ex:CS CL not CC}
Consider the LTSI with states $P,Q,R,S$ and transitions $t:P\tran a Q$, $u:P \tran b R$,
$t':R \tran {a'} S$ and $u':Q \tran {b'} S$, with empty independence relation.
This is essentially the same as Example~\ref{ex:WFnotCC},
except that we have disambiguated the transition labels, to reflect that
the four transitions
form four different events.
Then CC does not hold, but we claim that both CS$\indt$ and CL$\indt$ hold.

CS$\indt$: There are four possible cases to check, depending on the initial forward transition.
Consider first $t:P \tran a Q$ and some $r:Q \ptran\rho Q'$, $P' \tran a Q'$, where
$\cte(r,[t]) = 0$ and $(P,a,Q) \sqeqt (P',a,Q')$. 
Clearly $P' = P$ and $Q' = Q$.
To verify CS$\indt$ in this case, it is enough to show that $\cte(r,[u]) = \cte(r,[t']) = \cte(r,[u']) = 0$.
Since $r$ is a circuit, it enters each state as often as it leaves it.
Furthermore, since $\cte(r,[t]) = 0$, $r$ enters $Q$ from $P$ as often as it leaves $Q$ towards $P$.
Hence $r$ must enter $Q$ from $S$ as often as it leaves $Q$ towards $S$,
meaning that $\cte(r,[u']) = 0$.
We can similarly deduce that $\cte(r,[t']) = 0$ and $\cte(r,[u]) = 0$.
The remaining three cases with initial transitions $u$, $t'$ and $u'$ are similar to the case for $t$.

CL$\indt$: Again there are four cases to check, depending on the initial forward transition.
Consider first $t:P \tran a Q$ and some $r:Q \ptran\rho Q'$ where
$\cte(r,[t]) = 0$ and for all $t''$ in $r$ we have 
$\cte(r,[t'']) \leq 0$ (indeed, if $\cte(r,[t'']) > 0$ we would require $\rev{t} \ind t''$, which is false since the independence relation is empty, hence the condition for CL$\indt$ would hold trivially). However, if $\cte(r,[t'']) < 0$ then
there is $t'''$ in $r$ with $[t'''] = [\rev{t''}]$ (in this example actually $t''' = \rev{t''}$) and 
$\cte(r,[\rev{t''}]) > 0$, but, for the same reason as above, we cannot have $\cte(r,[\rev{t''}]) > 0$  since the independence relation is empty. Hence for each $t''$ we have $\cte(r,[t'']) = 0$, which implies $Q'=Q$, since the net rotation (cfr.\ Figure~\ref{fig:WFnotCC}) of each transition is zero, and so the net rotation of $r$ is zero. The thesis follows trivially.
The remaining three cases with initial transitions $u$, $t'$ and $u'$ are similar to the case for~$t$.\finex
\end{example}

The next axiom states that independence is fully determined by its restriction to coinitial transitions. It is related to axiom (E) of~\cite[page 325]{SNW96},
but here we allow reverse as well as forward transitions.
\begin{definition}{\bf Independence of events is coinitial (IEC)}\label{def:IEC}:
if $t_1 \ind t_2$ then $[t_1] \coind [t_2]$.
\end{definition}

Thanks to previous axioms, independence behaves well
w.r.t.~reversing.
\begin{definition}{\bf Reversing preserves independence (RPI)}\label{def:rpi}:
  if $t \ind t'$ then $\rev t \ind t'$.
\end{definition}
%
 \begin{proposition}\label{prop:RPI}
If an LTSI satisfies SP, PCI, IRE, IEC then it also satisfies RPI.
\end{proposition}
\begin{proof}
Suppose $t \ind u$.
We must show $\rev t \ind u$.
By IEC we have $t' \sqeqt t$, $u' \sqeqt u$ such that $t' \ind u'$ and $t',u'$ are coinitial.
By SP there is a diamond $t',u',t'',u''$ with $t' \sqeqt t''$, $u' \sqeqt u''$.
Then $\rev{t'} \ind u''$ using PCI.
Then $\rev t \sqeqt \rev{t'} \ind u'' \sqeqt u$
and so by IRE $\rev t \ind u$ as required.
\end{proof}
We can use IEC or IRE to show that transitions which are part of the same event cannot be independent.
\begin{definition}{\bf Event coherence (ECh)}\label{def:ECh}:
if $t \sqeqt t'$ then $t \notind t'$.
\end{definition}
\begin{proposition}\label{prop:ECh}
If a pre-reversible LTSI satisfies either IRE or IEC then it also satisfies ECh.
\end{proposition}
\begin{proof}
Assume for a contradiction that $t \sqeqt t'$ and $t \ind t'$.
First suppose that IRE holds.
We deduce $t \ind t$, contradicting irreflexivity of $\ind$.
Now suppose that IEC holds.
Then $[t] \coind [t']$, and so $[t] \coind [t]$, contradicting irreflexivity of $\coind$ (Proposition~\ref{prop:coind irref}).
\end{proof}


All the axioms that we have introduced so far are independent,
i.e.\ none is derivable from the remaining axioms.

The next example shows that IRE is not implied by other axioms.
\begin{example}\label{ex:CLG CSi}
Let $t:P \tran a Q$, $u:P \tran b R$,
$u':Q \tran b S$, $t':R \tran a S$,
with $t \ind u$, $\rev u \ind t'$, $\rev{t'} \ind \rev{u'}$, $u' \ind \rev t$, namely we have independence at all corners of the diamond.
Here we have two forward events, labelled with $a$ and $b$ respectively.
We have $t' \sqeqt t \ind u$ but not $t' \ind u$, so that IRE fails.
However axioms SP, BTI, WF, PCI and IEC hold.\finex
\end{example}
The next example shows that IEC is not implied by other axioms.
\begin{example}\label{ex:IRE1}
Let $t:P \tran a Q$, $u:R \tran b S$,
where all states are distinct,
and let $t \ind u$.
Then IEC fails;
however axioms SP, BTI, WF, PCI and IRE hold.\finex
\end{example}
The counterexample above remains valid also if $Q=R$, as shown below.
\begin{example}\label{ex:IRE2}
Let $t:P \tran a Q$, $u:Q \tran b S$,
and let $t \ind u$.
Then IEC fails;
however axioms SP, BTI, WF, PCI and IRE hold.\finex
\end{example}

We can now prove the independence result.
\begin{proposition}\label{prop:ind}
The axioms 
SP, BTI, WF, PCI, IRE and IEC are independent of each other.
\end{proposition}
\begin{proof}
For each of the six axioms we give an LTSI which satisfies the other five axioms but not the axiom itself.
In each case it is straightforward to check that the remaining axioms hold.

{\bf SP:} Let $t:P \tran a Q$ and $u:P \tran b R$ with $t \ind u$.

{\bf BTI:} Let $P \tran a R$ and $Q \tran b R$ with an empty independence relation
(Example~\ref{ex:notPL}).

{\bf WF:} Let $P_{i+1} \tran a P_i$ for $i = 0,1,\ldots$ with an empty independence relation.

{\bf PCI:} Let $t:P \tran a Q$, $u:P \tran b R$,
$u':Q \tran b S$, $t':R \tran a S$,
with $\rev{t'} \ind \rev{u'}$.

{\bf IRE:} See Example~\ref{ex:CLG CSi}.

{\bf IEC:} See Example~\ref{ex:IRE1} or Example~\ref{ex:IRE2}.
\end{proof}

\subsection{CS and CL via independent events}\label{sub:indev}

We now introduce a second version of causal safety and liveness,
which uses independence like CS$\indt$ and CL$\indt$,
but on events rather than on transitions. More precisely, we use coinitial independence $\coind$.

\begin{definition}\label{def:coind safe live}
Let $\mc L = (\Proc,\Lab,\tran{},\ind)$ be an LTSI. 
\begin{enumerate}
\item
We say that $\mc L$ is \emph{coinitially causally safe} (CS$\ci$) if whenever
$t_0:P \tran a Q$, $r:Q \ptran \rho R$, $\cte(r,[t_0]) = 0$ and $t_0\op:S \tran a R$
with $t_0 \sqeqt t_0\op$,
then $[\rev{t_0}] \coind e$ for all 
events $e$ such that $\cte(r,e) > 0$.
\item
We say that $\mc L$ is \emph{coinitially causally live} (CL$\ci$) if whenever
$t_0:P \tran a Q$, $r:Q \ptran \rho R$ and $\cte(r,[t_0]) = 0$ and
$[\rev{t_0}] \coind e$, for all 
events $e$ such that $\cte(r,e) > 0$,
then we have
$t_0\op:S \tran a R$ with $t_0 \sqeqt t_0\op$.
\end{enumerate}
\end{definition}
Note that in Definition~\ref{def:coind safe live} we operate at the level of events,
rather than at the level of transitions as in Definition~\ref{def:safe live}.
Also note that we could replace $[\rev{t_0}] \coind e$ by
$[t_0] \coind e$ using Lemma~\ref{lem:coind rev}.
We have used the former for compatibility with Definition~\ref{def:safe live}.
%
\begin{theorem}\label{thm:CS coind}
If an LTSI is pre-reversible then it satisfies CS$\ci$.
\end{theorem}
%
\begin{proof}
Suppose $t_0:P \tran a Q$, $r:Q \ptran \rho R$
and $t_0\op:S \tran a R$
with $t_0 \sqeqt t_0\op$.
By Lemma~\ref{lem:ladder} there is a path $s$ from $Q$ to $R$
such that for all $u$ in $s$ we have
$[t_0] \coind [u]$.
By CC, $r \ceqt s$.

Suppose that $e$ is an event and $\cte(r,e) > 0$.
Then $\cte(s,e) > 0$, thanks to Lemma~\ref{lemma:cccount}.
Hence there is $u$ in $s$ such that $[u] = e$. 
Since $[t_0] \coind [u]$,
also $[t_0] \coind e$.
Hence $[\rev{t_0}] \coind e$ using
Lemma~\ref{lem:coind rev}. 
\end{proof}

We now introduce a weaker version of axiom IRE (Definition~\ref{def:IRE}).
\begin{definition}{\bf Coinitial IRE (CIRE)}\label{def:CIRE}:
if $[t] \coind [u]$ and $t,u$ are coinitial then $t \ind u$.
\end{definition}
It is easy to see that IRE implies CIRE.
By considering Example~\ref{ex:CLG CSi}
we see that an LTSI can be pre-reversible and satisfy CIRE (and IEC) but not IRE.
Also, CIRE is not sufficient to ensure ECh (Definition~\ref{def:ECh}) holds, as shown by the next example.
\begin{example}\label{ex:CSi+RPI CLi}
Let $t:P \tran a Q$, $u:P \tran b R$,
$u':Q \tran b S$, $t':R \tran a S$.
We add independence between all pairs of distinct transitions
drawn from $t,u,t',u'$.
We furthermore add those independent pairs derived from closing under RPI.
We see that the LTSI is pre-reversible.  It satisfies CIRE and RPI,
but not ECh, since $t \sqeqt t'$ and also $t \ind t'$.
\finex
\end{example}

The next example shows that notions of CS/CL based on independence on transitions and on coinitial independence of events are not equivalent.
\begin{example}\label{ex:IC CLi}
Consider the LTSI in Figure~\ref{fig:IC CLi}.
\begin{figure}[!t]
\psfrag{a}{$a$}
\psfrag{b}{$b$}
\psfrag{c}{$c$}
\psfrag{P}{$P$}
\psfrag{Q}{$Q$}
\psfrag{R}{$R$}
\begin{center}
\epsfig{file=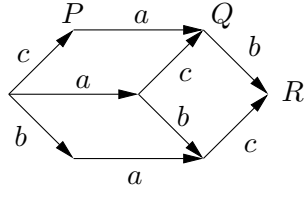, height=2.5cm}
\caption{
The LTSI in Example~\ref{ex:IC CLi}. 
}\label{fig:IC CLi}
\end{center}
\end{figure}
Independence is given by closing under BTI and PCI. Clearly WF and SP hold; hence the LTSI is pre-reversible and satisfies CS$\ci$.
There are three events, labelled $a,b,c$, which are all independent of each other.
Furthermore IEC holds, but not CIRE
(noting that the leftmost $b$ and $c$ transitions are coinitial but not independent, while the corresponding events are coinitially independent thanks to the rightmost square).
Also CL$\ci$ fails: consider $P \tran a Q \tran b R$,
where $a$ cannot be reversed at $R$ even though $[Q \tran{\rev{a}} P] \coind [Q \tran b R]$.
Differently from CS$\ci$, CS$\indt$ fails:
e.g., from the leftmost corner one can do $bac\rev b$, reversing $b$, but the inverse of the first $b$-transition is not independent with the $c$-transition.
Differently from CL$\ci$, CL$\indt$ holds:
the only state at which any event that has occurred cannot be
immediately reversed is $R$.
So we can restrict attention to instances of $P' \tran a Q'$, $r:Q' \ptran \rho R$.
Furthermore $r$ must finish with either $Q \tran b R$ or the $c$ transition to $R$.
These two transitions are not independent with any inverse $a$ transition.
Hence CL$\indt$ holds in these cases vacuously.\finex
\end{example}
\begin{proposition}\label{prop:CSindt RPI CIRE}
Let $\mc L$ be a pre-reversible LTSI.
If $\mc L$ satisfies CS$\indt$ and RPI then $\mc L$ also satisfies CIRE.
\end{proposition}
\begin{proof}
Assume that $\mc L$ satisfies CS$\indt$.
Suppose that $t,u$ are coinitial transitions such that $[t] \coind [u]$.
We must show that $t \ind u$.
We can suppose that at least one of $t$ and $u$ is forward;
otherwise we can obtain $t \ind u$ from BTI.
Without loss of generality, suppose that $t:P \tran a Q$ is forward.
Since $[t] \coind [u]$, there are coinitial $t':P' \tran a Q'$ and $u'$
such that $t \sqeqt t' \ind u' \sqeqt u$.
By SP we can complete a square containing $t',u'$ and two further transitions
$t'' \sqeqt t'$ and $u'' \sqeqt u'$ both with the same target~$R$.

By Lemma~\ref{lem:ladder} there is a path $s:Q \ptran \rho Q'$.
Let $r' = \rev t u \rev u t s$ (a path from $Q$ to $Q'$),
and consider the path
$r = r'u''$
from $Q$ to $R$.
We see that $\cte(r,[t])= 0$,
using Lemma~\ref{lem:cte zero} applied to $t,t''$ and~$r$.
Hence CS$\indt$ applies to $t$ together with $r$ and $t''$.
We deduce that
$\rev t \ind u_1$ for all $u_1$ in $r$ such that $\cte(r,[u_1]) > 0$.
We see that $\cte(tr',[u]) = 0$ using
Lemma~\ref{lem:cte zero} applied to $\rev u,\rev{u''}$ and $tr'$.
Noting that $\und{[t]} \neq \und{[u]}$
by 
Proposition~\ref{prop:coind und},
we obtain $\cte(r,[u]) = 1$
and so $\rev t \ind u$.
We deduce $t \ind u$ using RPI.
\end{proof}
We cannot omit the assumption of RPI in Proposition~\ref{prop:CSindt RPI CIRE},
in view of the following example.
\begin{example}\label{ex:halfcube mod}
Consider the `half cube' LTSI with transitions $a,b,c$ in Figure~\ref{fig:halfcube}.
\begin{figure}[!t]
\psfrag{a}{$a$}
\psfrag{b}{$b$}
\psfrag{c}{$c$}
\psfrag{P}{$P$}
\psfrag{Q}{$Q$}
\psfrag{Q'}{$Q'$}
\psfrag{R}{$R$}
\psfrag{S}{$S$}
\begin{center}
\epsfig{file=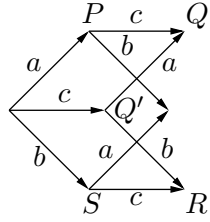, height=3cm}
\end{center}
\caption{LTSI of Examples~\ref{ex:halfcube} and~\ref{ex:halfcube mod}}\label{fig:halfcube}
\end{figure}
We add independence as given by BTI and PCI, and also between all pairs of transitions $t,u$ where at least one of $t,u$ is backward,
and $t \not\sqeqt u$, $t \not\sqeqt \rev u$. Clearly RPI does not hold. 
The LTSI is pre-reversible, and IEC holds.
CIRE does not hold; note that the $a$ and $b$-events are independent,
but after performing $c$ there are coinitial $a$ and $b$-transitions
which are not independent.
Both CL$\ci$ and CL$\indt$ hold: 
note that at any state, all events that have occurred can be reversed
immediately.
We have ensured that CS$\indt$ holds, since all independence deducible from CS$\indt$ must involve a backward transition $\rev{t_0}$ and a transition $u$ such that $t_0 \not\sqeqt u$ and
$t_0 \not\sqeqt \rev u$.\finex
\end{example}

We can characterise CIRE as being equivalent to coinitial transitions
with a common derivative process being independent.
\begin{proposition}\label{prop:char CIRE}
Let $\mc L$ be a pre-reversible LTSI.  The following are equivalent:
\begin{enumerate}
\item\label{item:CIRE}
$\mc L$ satisfies CIRE;
\item\label{item:CDI}
If $t:P \tran \alpha Q$, $r:Q \ptran \rho S$ and 
$u:P \tran \beta R$, $s:R \ptran \sigma S$ where 
$\und\alpha \neq \und\beta$ and
$\cte(r,[t]) = \cte(s,[u]) = 0$ then
$t \ind u$.
\end{enumerate}
\end{proposition}
\begin{proof}
Assume (\ref{item:CIRE}).
Let $t:P \tran \alpha Q$, $r:Q \ptran \rho S$ and 
$u:P \tran \beta R$, $s:R \ptran \sigma S$ where 
$\und\alpha \neq \und\beta$ and
$\cte(r,[t]) = \cte(s,[u]) = 0$.
We must show $t \ind u$.
Since the LTSI is pre-reversible, polychotomy holds for events $[t]$ and $[u]$
(Proposition~\ref{prop:poly}).
We can exclude $[t] = [u]$ since
$\und\alpha \neq \und\beta$.
There is a rooted path $r_0$ from some irreversible $I$ to $P$.
Since NRE holds (Proposition~\ref{prop:NRE}),
$\cte(r_0,[t]) = \cte(r_0,[u])$.
By considering the paths $r_0t $ and $r_0u$ we deduce that neither $[u] < [t]$ nor $[t] < [u]$ hold.
By CC applied to $tr$ and $us$ we see that $\cte(r,[u]) = 1$.
Hence $r_0tr$ is a rooted path with $\cte(r_0tr,[t]) = \cte(r_0tr,[u]) = 1$,
so that we can exclude $[t] \cf [u]$.
By polychotomy we conclude that $[t] \coind [u]$.
Then $t \ind u$ by CIRE.

Assume (\ref{item:CDI}).
Let $[t] \coind [u]$ where
$t:P \tran \alpha Q$ and $u:P \tran \beta R$ are coinitial.
We must show $t \ind u$.
First note that $\und\alpha \neq \und\beta$
by Proposition~\ref{prop:coind und}.
We have $t \sqeqt t' \ind u' \sqeqt u$ where
$t':P' \tran \alpha Q'$ and $u':P' \tran \beta R'$ are coinitial.
By SP we have
$t'':R' \tran \alpha S$ and $u'':Q' \tran \beta S$.
By Lemma~\ref{lem:ladder} we have
$r':Q \ptran\rho Q'$ such that for all $u_1$ in $r'$ we have $[t] \coind [u_1]$,
and $s':R \ptran\sigma R'$ such that for all $u_2$ in $s'$ we have $[u] \coind [u_2]$.
Let $r = r'u''$ and $s = s't''$.
We have $\cte(r,[t]) = \cte(s,[u]) = 0$ using Lemma~\ref{lem:cte zero}.
Hence $t \ind u$ as required, using the hypothesis.
\end{proof}

Notably, in the proof of (\ref{item:CIRE}) $\Rightarrow$ (\ref{item:CDI}),
CIRE is only used in the last step. Hence, the result could be rephrased by stating that any pre-reversible LTSI satisfies (\ref{item:CDI}),
with a conclusion of $[t] \coind [u]$ rather than $t \ind u$.

The independence result in Proposition~\ref{prop:ind} holds also if we replace IRE by CIRE.
\begin{proposition}\label{prop:ind CIRE}
The axioms 
SP, BTI, WF, PCI, CIRE and IEC are independent of each other.
\end{proposition}
\begin{proof}
For each of the six axioms we need to give an LTSI which satisfies the other five axioms but not the axiom itself.
Since IRE implies CIRE, for all axioms apart from CIRE we can reuse the examples given in the proof of Proposition~\ref{prop:ind}.
Example~\ref{ex:IC CLi} provides an LTSI where CIRE fails and the remaining five axioms hold.
\end{proof}

We can distinguish three mutually exclusive cases for CIRE
(Definition~\ref{def:CIRE}):
\begin{description}
\item
[forward case:] both transitions are forward;
\item
[backward-forward case:] one transition is backward, one is forward;
\item
[backward case:] both transitions are backward (implied by BTI).
\end{description}
The second case is particularly relevant for the characterisation of CL$\ci$; hence we state it as a separate axiom.
\begin{definition}\label{def:BFCIRE}
{\bf Backward-Forward CIRE (BFCIRE)}:
if $t:P \tran a Q$ and $u:Q \tran b R$ and $[\rev t] \coind [u]$ then $\rev t \ind u$.
\end{definition}
Thus BFCIRE is just CIRE specialised to the case where one of the
coinitial transitions is backward and one is forward. It has some similarity with one of the properties of transition systems with independence in \cite{NW95} and \cite[Definition 4.1]{SNW96}, and Sideways Diamond properties in \cite{PU07a,Aub22}. However, all of these properties state that if two consecutive forward transitions are independent then they are two sides of a commuting diamond.   

Analogously to what was done in Theorem~\ref{thm:CL} for CL$\indt$, we give below conditions for ensuring CL$\ci$.
Notably, here BFCIRE is necessary and sufficient,
while for CL$\indt$ we required IRE, which was sufficient but not necessary.

\begin{theorem}\label{thm:CL coind}
Let $\mc L$ be a pre-reversible LTSI.  Then the following are equivalent:
\begin{enumerate}
\item\label{item:L BFCIRE}
$\mc L$ satisfies BFCIRE;
\item\label{item:L CLci}
$\mc L$ satisfies CL$\ci$.
\end{enumerate}
\end{theorem}
\begin{proof}
Assume (\ref{item:L BFCIRE}).
  Suppose $t_0:P \tran a Q$, $r:Q \ptran \rho R$ and $\cte(r,[t_0]) =
  0$ and $[\rev{t_0}] \coind e$, for all $e$ such that $\cte(r,e) > 0$.  We
  have to show that there is $t_0\op:S \tran a R$ with $t_0 \sqeqt
  t_0\op$.

  Thanks to PL, there is $T$ such that $b: P
  \ptran{\rho_b} T$ and $f: T \ptran{\rho_f} R$, with $b$ backward and
  $f$ forward.
  By CC, $t_0r \ceqt bf$.
  Since $\cte(r,[t_0]) = 0$, thanks to
  Lemma~\ref{lemma:cccount} $\cte(bf,[t_0])=1$. As a consequence,
  there is a transition $t'_0:P' \tran a Q' \in [t_0]$ in
  $f$ (which is unique by Proposition~\ref{prop:NRE}).
  Let $f'$ be the portion of $f$ from $Q'$ to $R$.

  If we can show that $[\rev{t_0}] \coind [t'']$ for each transition $t''$ in $f'$,
  then the thesis will follow by commuting
  $t'_0$ with all such transitions using SP and BFCIRE.

  By Lemma~\ref{lem:ladder} there is a path $s$ from $Q$ to $Q'$ such that $[t_0] \ind [u]$  
  for all $u$ in $s$.
  By CC, $r \ceqt sf'$.
  Take any $t''$ in $f'$.
  By Lemma~\ref{lemma:cccount}, $\cte(r,[t'']) = \cte(s,[t'']) + \cte(f',[t''])$.
  If $\cte(s,[t'']) < 0$ then there is $u$ in $s$ such that $u \sqeqt \rev{t''}$.
  Now $[t_0] \coind [u]$, and so $[t_0] \coind [t'']$ using Lemma~\ref{lem:coind rev}.
  So suppose $\cte(s,[t'']) \geq 0$.
  Since $\cte(f',[t'']) > 0$, we have $\cte(r,[t'']) > 0$. 
  So there is $u$ in $r$ such that $u \sqeqt t''$,
  and by hypothesis $[\rev{t_0}] \coind [u]$, so that $[\rev{t_0}] \coind [t'']$.

Assume (\ref{item:L CLci}). 
Suppose that $t_0:P \tran a Q$ and $u:Q \tran b R$ and $[\rev {t_0}] \coind [u]$.
Clearly $\cte(u,[t_0]) = 0$.
By CL$\ci$ we have $t_0^{\dagger}:S \tran a R$ with $t_0 \sqeqt t_0^{\dagger}$.
Using BTI and SP we can complete a square starting with $\rev u$ and
$\rev{t_0^{\dagger}}$.
Using BLD this square must include $t_0$.
Using PCI we see that $\rev{t_0} \ind u$ as required.
\end{proof}
CL$\ci$ (and BFCIRE) do not imply CIRE,
as shown by Example~\ref{ex:halfcube mod}.
\begin{lemma}\label{lem:CSi BFCIRE}
Let a pre-reversible LTSI satisfy CS$\indt$.  Then it satisfies BFCIRE.
\end{lemma}
\begin{proof}
Suppose $t_0:P \tran a Q$ and $u:Q \tran b R$ and $[\rev {t_0}] \coind [u]$.
We must show that $\rev {t_0} \ind u$.

By Lemma~\ref{lem:coind rev} $[t_0] \coind [u]$ and so there are coinitial
$t_0':P' \tran a Q'$ and $u':P' \tran b R'$ with
$t_0 \sqeqt t'_0 \ind u' \sqeqt u$. 
Using SP we can complete a square with $t_0^{\dagger}:R' \tran a S'$ and $u'':Q' \tran b S'$.
By Lemma~\ref{lem:ladder} applied to $u$ and $u''$ we have a path $s$ from $R$ to $S'$.
Let $r = us$.
Then $\cte(r,[t_0]) = 0$ using Lemma~\ref{lem:cte zero}.
Also $\cte(s,[u]) = 0$ using Lemma~\ref{lem:cte zero},
so that $\cte(r,[u]) > 0$.
By CS$\indt$ applied to $t_0,t_0^{\dagger}$ and $r$
we deduce $\rev{t_0} \ind u$ as required.
\end{proof}

Perhaps surprisingly,
we can now relate safety with independence of transitions to liveness with independence of events.
\begin{proposition}\label{prop:CSi CLci}
Let a pre-reversible LTSI satisfy CS$\indt$.
Then it satisfies CL$\ci$.  
\end{proposition}
\begin{proof}
By Lemma~\ref{lem:CSi BFCIRE} and Theorem~\ref{thm:CL coind}.
\end{proof}

CL$\ci$ (and BFCIRE) do not imply CS$\indt$,
as shown by the next example.
\begin{example}\label{ex:halfcube}
Consider the `half cube' LTSI with transitions $a,b,c$ in Figure~\ref{fig:halfcube}.
We add independence as given by BTI and PCI.
The LTSI is pre-reversible.
As in Example~\ref{ex:halfcube mod}, CIRE does not hold
while both CL$\ci$ (hence BFCIRE) and CL$\indt$ hold.  
All pairs of independent transitions are coinitial.
CS$\indt$ however does not hold:
consider $t_0:P \tran c Q$, $r:Q \ptran {\rev a b} R$, $S \tran c R$---here we do not have $\rev {t_0} \ind (Q',b,R)$.\finex
\end{example}

\begin{proposition}\label{prop:correspondence}
Let $\mc L$ be a pre-reversible LTSI satisfying IEC.
If $\mc L$ satisfies CL$\ci$ then $\mc L$ satisfies CL$\indt$.
\end{proposition}
\begin{proof}
Immediate from the definitions.
\end{proof}

We next give an example where CC holds but not CS$\ci$ (and not PCI).
\begin{example}\label{ex:CC not CS}
  Consider the cube with transitions $a,b,c$ on the left in Figure~\ref{fig:diamonds},
where the forward direction is from left to right.
\begin{figure}[!t]
\psfrag{a}{$a$}
\psfrag{b}{$b$}
\psfrag{c}{$c$}
\psfrag{P_0}{$P_0$}
\psfrag{P_1}{$P_1$}
\psfrag{P_2}{$P_2$}
\psfrag{Q_0}{$Q_0$}
\psfrag{Q_1}{$Q_1$}
\psfrag{Q_2}{$Q_2$}
\psfrag{P}{$P$}
\psfrag{Q}{$Q$}
\begin{center}
\epsfig{file=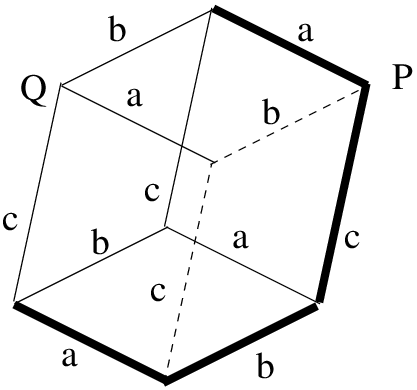, height=3.5cm}\qquad\qquad\qquad
\epsfig{file=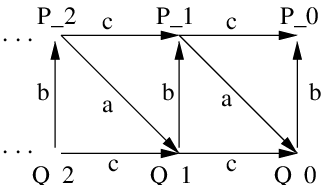, height=2.5cm}
\caption{
  The LTSIs in Examples~\ref{ex:CC not CS} and~\ref{ex:CS+CL not CC}.
  In the left-hand diagram, forward transitions are from left to right, and we use different line styles to make discussion of the diagram clearer.
}\label{fig:diamonds}
\end{center}
\end{figure}
We add independence as given by BTI.  So SP, BTI, WF hold, but not PCI.
Consider the bold path from the leftmost end:
we have an $a$-transition followed by a path $r = bc$ followed by $\rev a$.
For CS$\ci$ to hold, we want $\rev a$ to be the reverse of the same event as the first $a$.
They are connected by a ladder with sides $cb$.
We add independence for all corners on the two faces of the ladder ($ac$ and $ab$). Transitions
$\rev b$ and $\rev c$ at $P$ are independent (by BTI) so we obtain $\rev b\rev c \ceqt \rev c \rev b$, where 
$\rev b\rev c$ is dashed and $\rev c \rev b$ is bold. Since $\ceqt$ is closed under composition, we get
$bc \ceqt cb$.
However the bold $b$ is a different event from the event of the top $b$s since the bold-dashed $bc$ face does not have independence at each corner.
Therefore we do not get $[a] \coind [b]$ for the bold $a$ and bold $b$, and CS$\ci$ fails. However, we note that we do have $[a] \coind [b]$ for the bold $a$ and the dashed $b$ since $a$ and $b$ at $Q$ are independent.
\end{example}
We next give an example where CS$\ci$ and CL$\ci$ hold 
but not CC.



\begin{example}\label{ex:CS+CL not CC}
Consider the LTSI with
$Q_i \tran b P_i$, $P_{i+1} \tran c P_i$, $Q_{i+1} \tran c Q_i$,
$P_{i+1} \tran a Q_i$ for $i = 0,1,\ldots$. This is shown on the right in Figure~\ref{fig:diamonds}.
Clearly WF does not hold.
We add coinitial independence to make BTI and PCI hold.
Then also SP and CIRE hold.
However, CC fails since, for example
$P_1 \tran a Q_0 \tran b P_0$ and $P_1 \tran c P_0$ are coinitial and cofinal but not causally equivalent.
Note that there are just three events $a,b,c$ with $a \coind c$,
$b \coind c$ but not $a \coind b$.
CS$\ci$ and CL$\ci$ hold. Indeed, $c$ is independent from every other action, and it can always be undone, while $a$ and $b$ are independent from $c$ only and they can be undone after any path composed by $c$ and no others.
In more detail, if we have a path $a r \rev a$ with $\cte(r,a) = 0$ then $\cte(r,b) = 0$,
and if we have a path $b r \rev b$ with $\cte(r,b) = 0$ then $\cte(r,a) = 0$.\finex
\end{example}

The independence result in Proposition~\ref{prop:ind CIRE} holds also if we replace CIRE by BFCIRE.
\begin{proposition}\label{prop:ind BFCIRE}
The axioms 
SP, BTI, WF, PCI, BFCIRE and IEC are independent of each other.
\end{proposition}
\begin{proof}
%
For each of the six axioms we need to give an LTSI which satisfies the other five axioms but not the axiom itself.
Since CIRE implies BFCIRE, for all axioms apart from BFCIRE we can reuse the examples given in the proofs of Proposition~\ref{prop:ind CIRE}
(and of Proposition~\ref{prop:ind}).
Example~\ref{ex:IC CLi} provides an LTSI where BFCIRE (equivalent to CL$\ci$) fails and the remaining five axioms hold.
\end{proof}


\subsection{CS and CL via ordering of forward events}\label{sub:ord}
We now give definitions of causal safety and causal liveness 
using ordering on forward events.
To this end, we exploit the causality relation $\leq$ on such events (see Definition~\ref{def:ordering}).
\begin{definition}\label{def:safe live <}
Let $\mc L = (\Proc,\Lab,\tran{},\ind)$ be an LTSI.
\begin{enumerate}
\item
We say that $\mc L$ is \emph{ordered causally safe (CS$_<$)} if whenever
$t_0:P \tran a Q$, $r:Q \ptran \rho R$, $\cte(r,[t_0]) = 0$ and $t_0\op:S \tran a R$
with $t_0 \sqeqt t_0\op$,
then $[t_0] \not < e'$ for all forward events $e'$ such that $\cte(r,e') > 0$.
\item
We say that $\mc L$ is \emph{ordered causally live (CL$_<$)} if whenever
$t_0:P \tran a Q$, $r:Q \ptran \rho R$ and $\cte(r,[t_0]) = 0$ and
$[t_0] \not < e'$ for all forward events $e'$ such that $\cte(r,e') > 0$
then we have
$t_0\op:S \tran a R$ with $t_0 \sqeqt t_0\op$. 
\end{enumerate}
\end{definition}

The only difference between CS$_<$ and CS$\indt$
(Definition~\ref{def:safe live})
is that the former ensures $[t_0] \not < [t]$ instead of $\rev{t_0} \ind t$ for all transitions $t$ such that $[t]$ has a positive number of occurrences in $r$. Similarly for CL. Notably, we do not require  $[\rev{t_0}] \not < [t]$ since $<$ is defined on forward events and $t_0$ is forward.

It may seem that the definition above does not take into account backward events that may occur in $r$, but the next lemma shows that such events are necessarily independent from $[t_0]$.
This allows us to connect ordered safety and liveness
with 
safety and liveness based on independence of events.
\begin{lemma}\label{lem:coind <}
Suppose that an LTSI is pre-reversible.
Suppose $t_0:P \tran a Q$, $e = [t_0]$, $r:Q \ptran \rho R$ and $\cte(r,e) = 0$.
Let $e'$ be a forward event:
\begin{enumerate}
\item\label{item:greater}
if $\cte(r,e') > 0$ then exactly one of $e \coind e'$ and $e < e'$ holds;
\item\label{item:less}
if $\cte(r,e') < 0$ then $e \coind e'$.
\end{enumerate}
\end{lemma}
\begin{proof}
We know that polychotomy holds by Proposition~\ref{prop:poly}.
Also NRE holds by Proposition~\ref{prop:NRE}.
Suppose $t_0:P \tran a Q$, $e = [t_0]$, $r:Q \ptran \rho R$ and $\cte(r,e) = 0$ and
$\cte(r,e') \neq 0$ where $e'$ is a forward event.
We first note that $e \neq e'$, since $\cte(r,e) = 0$ and $\cte(r,e') \neq 0$.
By WF, there is a rooted path $s$ from some irreversible $I$ to $P$.
\begin{enumerate}
\item
Suppose first that $\cte(r,e') > 0$.
Since $\cte(st_0r,e) > 0$ and $\cte(st_0r,e') > 0$ we do not have $e \cf e'$.
Furthermore, if $e' < e$ then we must have $\cte(s,e') > 0$,
so that $\cte(st_0r,e') > 1$, contradicting NRE.
Then the result follows by polychotomy.

\item
Now suppose that $\cte(r,e') < 0$.
By Proposition~\ref{prop:regeqzero} we must have $\cte(s,e') > 0$.
We deduce that $e \not< e'$.
Since $\cte(st_0,e) > 0$ and $\cte(st_0,e') > 0$ we do not have $e \cf e'$.
Furthermore $\cte(st_0r,e) > 0$ and $\cte(st_0r,e') = 0$ (since $\cte(st_0,e') = 1$ combining $\cte(st_0,e') > 0$ shown above and NRE).
Hence $e' \not< e$.
By polychotomy, $e \coind e'$.
\qedhere
\end{enumerate}
\end{proof}

\begin{proposition}\label{prop:CS CL coind <}
Suppose that an LTSI $\mc L$ is pre-reversible.
Then
\begin{enumerate}
\item
$\mc L$ satisfies CS$_<$.
\item
$\mc L$ satisfies CL$\ci$ iff $\mc L$ satisfies CL$_<$.
\end{enumerate}
\end{proposition}
\begin{proof}
\begin{enumerate}
\item
We know CS$\ci$ holds by Theorem~\ref{thm:CS coind}.
Assume that $t_0:P \tran a Q$, $e = [t_0]$, $r:Q \ptran \rho R$, $\cte(r,e) = 0$ and $t_0\op:S \tran a R$
with $t_0 \sqeqt t_0\op$.
Take any forward $e'$ such that $\cte(r,e') > 0$.
By Lemma~\ref{lem:coind <} we know that exactly one of $e \coind e'$ or $e < e'$ holds.
By CS$\ci$ we have $e \coind e'$, and therefore $e \not< e'$ as required.

\item
  Suppose that CL$\ci$ holds.
Assume that
$P \tran a Q$, $e = [t_0]$, $r:Q \ptran \rho R$ and $\cte(r,e) = 0$ and
$e \not < e'$ for all 
forward $e'$ such that $\cte(r,e') > 0$.
Let event $e'$ be such that $\cte(r,e') > 0$.
Suppose first that $e'$ is forward.
By assumption $e \not < e'$.
So by Lemma~\ref{lem:coind <}(\ref{item:greater}) we obtain $e \coind e'$.
Suppose instead that $e'$ is reverse, so that $\rev {e'}$ is forward,
and $\cte(r,\rev{e'}) < 0$.
By Lemma~\ref{lem:coind <}(\ref{item:less}) we obtain $e \coind \rev{e'}$,
and hence $e \coind e'$ using Lemma~\ref{lem:coind rev}.
We deduce that $e \coind e'$ for all $e'$ such that $\cte(r,e') > 0$.
Hence by CL$\ci$ we have
$t_0\op:S \tran a R$ with $t_0 \sqeqt t_0\op$.

Conversely, suppose that CL$_<$ holds.
Assume that
$P \tran a Q$, $e = [t_0]$, $r:Q \ptran \rho R$ and $\cte(r,e) = 0$ and
$e \coind e'$ for all 
$e'$ such that $\cte(r,e') > 0$.
By Lemma~\ref{lem:coind <}(\ref{item:greater}) we know that $e \not < e'$ for all forward $e'$ such that $\cte(r,e') > 0$.
Hence by CL$_<$ we have
$t_0\op:S \tran a R$ with $t_0 \sqeqt t_0\op$.
\qedhere
\end{enumerate}
\end{proof}

\subsection{Implications between the different formalisations of CS/CL}\label{sub:comparing}

\begin{figure}[!t]
\psfrag{IRE}{IRE}
\psfrag{CIRE}{CIRE}
\psfrag{CSi}{CS$\indt$}
\psfrag{CLi}{CL$\indt$}
\psfrag{CLci<=>CL<}{BFCIRE $\Leftrightarrow$ CL$\ci \Leftrightarrow$ CL$_<$}
\psfrag{pre-reversible, CSci, CS<}{pre-reversible, CS$\ci$, CS$_<$}
\begin{center}
\epsfig{file=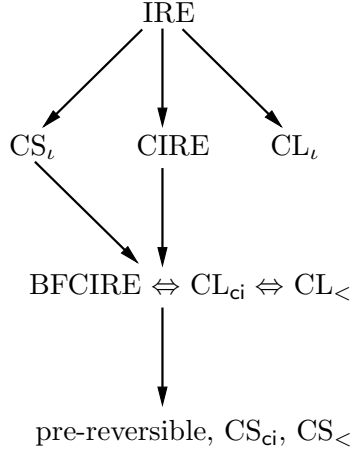, height=6.0cm}
\caption{
  Implications between causal safely and causal liveness properties and some of the related axioms, all assuming pre-reversibility.
  Note that both CS$\ci$ and CS$_<$ are implied by pre-reversibility.
}\label{fig:simpleCSCL}
\end{center}
\end{figure}

We have introduced three different formalisations of causal safety and liveness.  The implications between them, assuming pre-reversibility holds, 
are shown in Figure~\ref{fig:simpleCSCL}.

As can be seen in Table~\ref{t:list}, only two causal safety properties, namely CS$\ci$ and CS$_<$, hold for
pre-reversible LTSIs.  The causal liveness versions of these properties, namely CL$\ci$ and CL$_<$, additionally require BFCIRE. 
Actually, BFCIRE is equivalent to both CL$\ci$ and CL$_<$. The last two properties,
CS$\indt$ and CL$\indt$, which are defined over general independence of transitions, require  IRE. No other implications hold beyond those shown. Counterexamples for lack of other implications
in Figure~\ref{fig:simpleCSCL} are pointed to in Figure~\ref{fig:diagprerev1}.

We postpone discussion of which particular version of CS or CL is most relevant in a specific setting until Section~\ref{subsec:comparison}, after we have introduced some structural axioms to better relate them.

\section{Structured notions of independence}\label{sec:coinitial}
In this section we consider two structured notions of independence, namely independence defined on coinitial transitions only and independence determined by labels only.
To this end, we introduce `structural axioms' in Definitions~\ref{def:coinitial LTSI},~\ref{def:CLG} and~\ref{def:LG}.  These have a different status from the axioms already introduced: rather than expressing fundamental properties that are desirable in LTSIs,
they are properties that hold in various reversible formalisms 
(as we shall see in Section~\ref{sec:casestudies}), are easy to verify,
and can be used to derive other
axioms in a generic fashion.

\subsection{Coinitial independence}
In this section we discuss coinitial LTSIs, defined as follows,
and their relationship with LTSIs in general.

\begin{definition}\label{def:coinitial LTSI}
  {\bf Independence is coinitial (IC)}:
  for all transitions $t,u$,
if $t \ind u$ then $t$ and $u$ are coinitial.
\end{definition}

We say that an LTSI $\mc L$ is coinitial if it satisfies IC.
We also say that its independence relation $\ind$ is coinitial.

Coinitial independence is of interest since in many cases it is easier
to define independence only on coinitial transitions. Indeed,
coinitial independence arises, e.g., from the notions of concurrency
in~\cite[Definition 7]{DK04} for RCCS and in~\cite[Definition 5]{LaneseNPV18} for Core Erlang.

The next example satisfies IC and all and only the properties in Figure~\ref{fig:diagprerev1} implied by it. In particular, it shows that IC does not imply CL$\indt$, CL$\ci$, or
CS$\indt$ (this last follows from Proposition~\ref{prop:CSi CLci}).
\begin{example}\label{ex:IC}
Consider the LTSI in Figure~\ref{fig:IC}.
\begin{figure}[!t]
\psfrag{a}{$a$}
\psfrag{b}{$b$}
\psfrag{c}{$c$}
\psfrag{P}{$P$}
\psfrag{Q}{$Q$}
\psfrag{R}{$R$}
\psfrag{S}{$S$}
\begin{center}
\epsfig{file=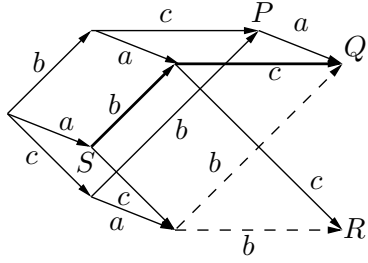, height=3.50cm}
\caption{
The LTSI in Example~\ref{ex:IC}.
}\label{fig:IC}
\end{center}
\end{figure}
Independence between transitions is generated by applying BTI and PCI and, as such, is coinitial as in Definition~\ref{def:coinitial LTSI}.
Moreover, the LTSI is pre-reversible.
There are three events, which we denote by $e_a,e_b,e_c$,
with labels $a,b,c$, respectively. 
CL$\indt$ fails:
let $t: P \tran a Q$ and let $r$ from $Q$ to $R$ be $\rev b b$ (dashed transitions).
We have $\cte(r,e_b) = 0$;
however $a$ cannot be reversed at $R$, as CL$\indt$ would yield.
Also CS$\indt$ fails:
let $t: P \tran a Q$ and let $r'$ be $\rev c\, \rev b$ from $Q$ to $S$ (bold transitions).
After $r'$, $\rev a$ is possible.
However $\rev t$ is not independent with the $\rev b$ transition,
as CS$\indt$ would yield.
Also CL$\ci$ fails:
let $t_0$ be the $a$ following the leftmost $b$, and let $r''$ be the $c$
transition with target $R$.  We have $e_a \coind e_c$.
However $a$ cannot be reversed at $R$, as CL$\ci$ would yield.
\finex
\end{example}

Coinitial independence is inconsistent with the axiom IRE,
showing that IRE is only appropriate for the setting of general,
rather than coinitial independence:
\begin{proposition}\label{prop:IC IRE}
Let a pre-reversible LTSI have a non-empty independence relation,
and satisfy IC.  Then IRE does not hold.
\end{proposition}
\begin{proof}
Suppose for a contradiction that IRE holds.
Since the independence relation is non-empty and IC holds,
we have $t \ind u$ with $t,u$ coinitial.
By SP and PCI we can complete a diamond with $t'\sqeqt t$, $u' \sqeqt u$.
Since $t' \sqeqt t \ind u$ we deduce by IRE that $t' \ind u$.
However $t'$ and $u$ are not coinitial,
contradicting IC.
\end{proof}

We define
a mapping~$\mathrel{c}$ restricting general independence to coinitial transitions and
a mapping~$\mathrel{g}$ extending independence along events.
\begin{definition}\label{def:gen coinit}
  Given an LTSI 
  $\mc L$, define
$t \mathrel{g(\ind)} u$ iff \mbox{$t \sqeqt t' \ind u' \sqeqt u$}
for some $t'$, $u'$.
Furthermore, define $t \mathrel{c(\ind)} u$
iff $t \ind u$ and $t$, $u$ are coinitial.
\end{definition}

  We extend $\mathrel{c}$ and $\mathrel{g}$ to LTSIs $(\Proc,\Lab,\tran{},\ind)$:
  they behave as the identity of the first three components, and as expected on the fourth. Similarly, we write $\mathrel{c(\sqeqt)}$ and $\mathrel{g(\sqeqt)}$ for the equivalence relations in $\mathrel{c(\mc L)}$ and $\mathrel{g(\mc L)}$, respectively.
  
We now show that $c$ and $g$ play well with events.
\begin{lemma}
 Given an LTSI $\mc L$, ${\sqeqt} = {c(\sqeqt)}$.
\end{lemma}
\begin{proof}
Follows by noticing that the definition of event only exploits independence on coinitial transitions.
\end{proof}

\begin{lemma}
 Given an LTSI $\mc L$, $t \sqeqt u$ implies $t \mathrel{g(\sqeqt)} u$.
\end{lemma}
\begin{proof}
By definition of $\sqeqt$, noticing that ${\ind} \subseteq {g(\ind)}$.
\end{proof}

\begin{lemma}
 Given a pre-reversible LTSI $\mc L$, $t \mathrel{g(\sqeqt)} u$ implies $t \sqeqt u$.
\end{lemma}
\begin{proof}
By definition of $\sqeqt$, we have $t \, g(\sqeqt) \, u$ if there is a chain of commuting squares connecting $t$ and $u$. Thanks to ID (which holds in pre-reversible LTSIs) all such squares are commuting squares in $\mc L$, hence $t \sqeqt u$ as desired.
\end{proof}

We can now study the impact of $c$ and $g$ on the axioms satisfied by the LTSI to which they are applied.
%
\begin{proposition}\label{prop:gen coinit}
Let $\mc L = (\Proc,\Lab,\tran{},\ind)$ be a pre-reversible LTSI. 
\begin{enumerate}
\item
if $\mc L$ is coinitial and satisfies CIRE then $c(g(\ind)) = {\ind}$;
\item
if $\mc L$ satisfies IRE and IEC then $g(c(\ind)) = {\ind}$;
\item\label{item:gprop}
If $\mc L$ is coinitial and satisfies CIRE then
$g(\mc L)$
is a pre-reversible LTSI and satisfies IRE and IEC.
\item
if $\mc L$ satisfies IRE then
$c(\mc L)$
is a pre-reversible coinitial LTSI and satisfies CIRE.
\end{enumerate}
\end{proposition}
\begin{proof}
\begin{enumerate}
\item
Clearly ${\ind} \subseteq c(g(\ind))$.
For the converse,
suppose $t \sqeqt t' \ind u' \sqeqt u$ and $t',u'$ are coinitial and $t,u$ are coinitial.
Then $t \ind u$ by CIRE.
\item
Suppose $t \ind u$.
By IEC we have $t \sqeqt t' \ind u' \sqeqt u$ with $t',u'$ coinitial.
Hence $t \mathrel{g(c(\ind))} u$.
Conversely, suppose $t \mathrel{g(c(\ind))} u$.
Then $t \ind u$ by IRE.
\item
Suppose $t \mathrel{g(\ind)} u$ and $t,u$ are coinitial.
Then by CIRE $t \ind u$.  So we can use SP for $\ind$ to complete the diamond.
Hence SP holds for $\mc L'$.

Clearly PCI holds for
$g(\mc L)$
since $g(\ind)$ and $\ind$ agree on coinitial transitions by CIRE.

For IRE, suppose $t' \sqeqt t \mathrel{g(\ind)} u \sqeqt u'$.  Then clearly $t' \mathrel{g(\ind)} u'$.

Finally, for IEC suppose $t \mathrel{g(\ind)} u$.  Then
$t \sqeqt t' \ind u' \sqeqt u$ with $t',u'$ coinitial, which is exactly what is needed
for IEC.
\item
Immediate.
\qedhere
\end{enumerate}
\end{proof}
Thanks to Proposition~\ref{prop:gen coinit}, we can extend
a coinitial pre-reversible LTSI satisfying CIRE in a canonical way to
a pre-reversible LTSI satisfying IRE and IEC.

Note that $g(\mc L)$ satisfies IRE (and hence ECh) by construction, since $t \mathrel{g(\ind)} u \sqeqt t'$ implies $t \mathrel{g(\ind)} t'$. Conditions in Proposition~\ref{prop:gen coinit}, item~(\ref{item:gprop}) are only needed for the other properties.


\subsection{Label-generated independence}
In some reversible calculi (such as RCCS) independence of coinitial transitions is defined purely
by reference to the labels.

\begin{definition}\label{def:CLG}
  {\bf Coinitial label-generated (CLG)}: if there is an irreflexive binary relation $I$ on $\Lab$, called a \emph{generator relation}, such that for any
transitions $t:P \tran\alpha Q$ and $u:P \tran\beta R$
we have $t \ind u$ iff $t$ and $u$ are coinitial and $I(a,b)$, where $a$ and $b$ are the underlying labels
$a = \und\alpha$, $b = \und\beta$.
\end{definition}

If this is the case then 
the axioms IC, PCI and CIRE hold by construction.
\begin{proposition}\label{prop:CLG}
If an LTSI is CLG then it satisfies IC, PCI and CIRE.
\end{proposition}
\begin{proof}
Straightforward, noting for PCI and CIRE that labels on opposite sides of a diamond of transitions must be equal.
\end{proof}

Note that $I$ must be irreflexive, since $\ind$ is irreflexive by definition. Even more, we already have seen that for a pre-reversible LTSI there cannot be independent coinitial transitions $t$, $u$ with the same underlying label (as a consequence of Lemma~\ref{lemma:revnotind} and BLD).

\begin{definition}\label{def:LG}
  {\bf Label-generated (LG)}: if there is an irreflexive binary relation $I$ on $\Lab$, called a \emph{generator relation}, such that for any
transitions $t:P \tran\alpha Q$ and $u:R \tran\beta S$
we have $t \ind u$ iff $I(a,b)$, where $a$ and $b$ are the underlying labels
$a = \und\alpha$, $b = \und\beta$.
\end{definition}
\begin{proposition}\label{prop:LG}
If an LTSI is LG then it satisfies PCI, IRE and RPI.
\end{proposition}
\begin{proof}
Straightforward.
\end{proof}
Note that LG does not imply IEC, in view of the following example.
\begin{example}\label{ex:LG}
Consider the LTSI with two transitions $t:P \tran a Q$ and $u:R \tran b S$, where all states are distinct (as in Example~\ref{ex:IRE1}) and $a \neq b$.
Let independence be generated by the relation $I = \{(a,b)\}$.
Then LG holds, but not IEC, since $t \ind u$ but not $[t] \coind [u]$.\finex
\end{example}

However, LG is compatible with IEC, in view of the following example.
\begin{example}\label{ex:LG+IEC}
Let $t:P \tran a Q$, $u:P \tran b R$,
$u':Q \tran b S$, $t':R \tran a S$,
where all states are distinct and $a \neq b$.
Let independence be generated by the relation $I = \{(a,b)\}$.
Then both LG and IEC hold.
However IC fails.
\finex
\end{example}

All the axioms and properties we have considered in the previous sections are closed under
disjoint unions of LTSIs, defined as follows.
\begin{definition}[Disjoint union of LTSIs]
Let $(\Proc_1,\Lab_1,\tran{}_1,\ind_1)$ and $(\Proc_2,\Lab_2,\tran{}_2,\ind_2)$ be LTSIs. Their disjoint union is $(\Proc_1 \cup \Proc_2,\Lab_1 \cup \Lab_2,\tran{}_1\cup\tran{}_2,\ind_1\cup\ind_2)$ provided that $\Proc_1 \cap \Proc_2 = \emptyset$, and undefined otherwise.
\end{definition}
However LG and CLG are not necessarily closed under disjoint unions of LTSIs,
in view of the following examples.
\begin{example}\label{ex:IRE+IEC}
Take the disjoint union of the LTSI of Example~\ref{ex:LG+IEC}
together with a further transition
$T \tran a U$ with an empty generator relation
(this component satisfies LG).
Then LG fails; however IEC and IRE still hold.
\finex
\end{example}
\begin{example}\label{ex:IC+CIRE}
Take the disjoint union of the LTSI of
Example~\ref{ex:prerev not CSi}
(which satisfies CLG)
together with further transitions $T \tran a U$ and $T \tran b V$
with an empty generator relation
(this component satisfies CLG).
Then CLG fails; however IC and CIRE still hold.
\finex
\end{example}

The mapping $g$ converts an LTSI satisfying CLG into one satisfying LG+IEC.
The mapping $c$ converts an LTSI satisfying LG into one satisfying CLG.
Note that there is an alternative way to convert an LTSI satisfying CLG into one satisfying LG:
simply use the relation $I$ applied to any pair of transitions.
This will in general create more independent transitions than using $g$,
and so the result may not satisfy IEC.

%
\subsection{Relating different forms of CS/CL}\label{subsec:comparison}
We now  discuss the relationships between different forms of CS/CL 
and consider which ones to work with in particular reversible settings. 
The starting point is how independence is or can be defined in such settings,
and whether it is general or coinitial. We explain how structural axioms and results of this section, 
together with our axioms, can be used to arrive at the most appropriate causal safety and liveness 
properties for such reversible settings. 

We can sometimes move between
LTSIs satisfying CS$\ci$ and CL$\ci$ (or equivalently  CS$_<$ and CL$_<$), all defined in terms of coinitial independence, and LTSIs satisfying CS$\indt$ and
CL$\indt$, which are based on general independence, using mappings $c$ and $g$.
Thus, if we have a coinitial pre-reversible LTSI $\mc L$ satisfying CIRE then CS$\ci$ and CL$\ci$
hold (using Theorems~\ref{thm:CS coind} and~\ref{thm:CL coind}, respectively).  
The LTSI $g(\mc L)$ is pre-reversible and satisfies IRE and IEC 
by Proposition~\ref{prop:gen coinit}. This will satisfy CS$\indt$ and CL$\indt$ as a result of 
applying Theorems~\ref{thm:CS} and~\ref{thm:CL}, respectively.
It will also satisfy CS$\ci$ and CL$\ci$. 
Conversely, if we have a general pre-reversible LTSI $\mc L'$ satisfying IRE then CS$\indt$ and CL$\indt$
hold by Theorems~\ref{thm:CS} and~\ref{thm:CL}, respectively. The LTSI $c(\mc L')$ 
is a coinitial pre-reversible LTSI satisfying CIRE.
This will satisfy CS$\ci$  and CL$\ci$.

Intuitively, one can think of coinitial independence as a compact way
of representing general independence (provided that this is
well-behaved, in that it satisfies IRE and IEC), and $c$ and $g$ as
ways of moving between the two representations (Proposition~\ref{prop:gen coinit}). 
CS$\indt$ and CL$\indt$ work on the general
representation only, since they check independence between transitions that
may be far apart. The other two forms of CS/CL can instead work with
both the representations, and they are equivalent (Figure~\ref{fig:simpleCSCL}). 
Moreover, once we have LTSI with general independence we can work immediately with 
CS$\indt$ and CL$\indt$. On the other hand, when independence is coinitial, we
need to instantiate the notion of event, and understand whether events are causally dependent or coinitial independent,
before we can use the other two notions of CS/CL.
The choice between CS$_<$/CL$_<$ and CS$\ci$/CL$\ci$ depends on whether independence 
or ordering is more easily or naturally defined on events. 

In some process calculi and programming languages, as can be seen in the next section,  
independence can be defined in terms of transition labels, which gives us structural axioms 
CLG and LG. So, to show CS/CL we tend to show CLG (RCCS, CCSK, HO$\pi$, Erlang) 
or we prove CIRE (R$\pi$, reversible occurrence nets) and then use $g$. 
Alternatively, we show LG ($\pi$IH).

Note that whether or not CLG/LG can be applied to a reversible formalism may depend on the level of abstraction adopted in the transition labels.

\section{Case Studies}\label{sec:casestudies}
We look at whether our axioms hold in various reversible formalisms.
Given that we consider a high number of formalisms, we do not provide full background on them, but refer for it to the original papers. Also, we sometimes repeat similar observations for different formalisms, so to make it possible to browse them out of order, to find information on a specific formalism of interest. 
Remarkably, all the works below provide proofs of the Loop Lemma.

\subsection{Reversible CCS (RCCS)}\label{sec:rccs}
We consider here the semantics of RCCS 
in~\cite{DK04}, and restrict the attention to coherent processes~\cite[Definition 2]{DK04}. 
In RCCS, transitions $P \tran {\mu:\zeta} Q$ and $P \tran {\mu':\zeta'} Q'$
are concurrent if $\mu \cap \mu' = \emptyset$ \cite[Definition~7]{DK04}.
This allows us to define coinitial independence as
$t \ind u$ iff $t$ and $u$ are concurrent.
We now argue that the resulting coinitial LTSI is pre-reversible and
also satisfies CIRE. SP was shown in~\cite[Lemma 8]{DK04}.
BTI was shown in the proof of~\cite[Lemma 10]{DK04}.
WF is straightforward, noting that backward transitions decrease memory size.
Hence, we obtain a very much simplified proof of CC.
For PCI and CIRE we note that CLG holds
and thus Proposition~\ref{prop:CLG} applies.
Therefore CS$\ci$ and CL$\ci$ hold.
Using Proposition~\ref{prop:gen coinit},
we can get an LTSI with general independence satisfying IRE and IEC,
and therefore CS$\indt$ and CL$\indt$.
This is the first time these causal properties have been proved for RCCS.

\subsection{CCS with Communication Keys (CCSK)}
The first notion of independence for CCSK~\cite{PU07} was given in~\cite{Aub22}. It is based on the proved transition system approach where transition labels
contain information about derivation of transitions. This information can be used to work out whether transitions are in conflict, causally dependent, or concurrent. Two forms of independence are defined in~\cite{Aub22}: general independence (called composable concurrency) and coinitial independence (called coinitial concurrency). CC is then obtained using our axiomatic approach (following~\cite{LanesePU20}, the conference version of the present paper) by showing SP \cite[Theorem~3]{Aub22}, BTI \cite[Lemma~6]{Aub22} and WF \cite[Lemma~7]{Aub22}.

Since coinitial independence is defined on labels, we can
deduce that the 
LTSI is CLG. Hence, by Proposition~\ref{prop:CLG}, PCI and CIRE hold. This allows us to obtain CS$\ci$ and CL$\ci$.
Using Proposition~\ref{prop:gen coinit},
we can get an LTSI with general independence which satisfies IRE and IEC, which gives us CS$\indt$ and CL$\indt$ as well. 
As for RCCS, this is the first time such causal properties have been 
proved for CCSK.

\subsection{Higher-Order $\pi$-calculus (HO$\pi$)}\label{sec:hopi}
We consider here the uncontrolled reversible semantics for HO$\pi$~\cite{LaneseMS16}. 
We restrict our attention to reachable
processes, called there consistent.
The semantics is a reduction semantics; hence there are no labels (or, equivalently, all
the labels coincide). To have more informative labels we
can consider the transitions defined in~\cite[Section~3.1]{LaneseMS16},
where labels contain the memory created or consumed by the transition
  (they also contain a flag distinguishing backward from forward transitions, but this plays no role in the definition of the concurrency relation discussed below, hence we can safely drop it). 
The notion of independence would be given by the concurrency relation on coinitial
transitions~\cite[Definition 9]{LaneseMS16}.
All pre-reversible LTSI axioms hold, as well as CIRE. 
Specifically, SP is proved in~\cite[Lemma 9]{LaneseMS16}. BTI holds since distinct memories have disjoint
sets of keys~\cite[Definition 3 and Lemma 3]{LaneseMS16} and by
the definition of concurrency~\cite[Definition 9]{LaneseMS16}.
WF holds as each backward step consumes a memory, which are a finite number to start with.
Finally, PCI and CIRE hold since CLG holds for the LTSI with annotated labels
and using our Proposition~\ref{prop:CLG}.
\Comment{
\begin{description}
\item[SP:] proved in~\cite[Lemma 9]{LaneseMS16};
\item[BTI:] since distinct memories have disjoint
  sets of keys~\cite[Definition 3 and Lemma 3]{LaneseMS16} and by 
  the definition of concurrency~\cite[Definition 9]{LaneseMS16};
\item[WF:] since each backward step consumes a memory;
\item[PCI, CIRE:] since the notion of concurrency is defined on the
  annotated labels and using our Proposition~\ref{prop:underlying}. 
\end{description}
%
}

As a result we obtain a very much simplified proof of CC.
Moreover, using PCI and CIRE, we get the CS$\ci$ and CL$\ci$ safety and liveness properties and, 
applying mapping $g$ from Section~\ref{sec:coinitial}, we get a general 
pre-reversible LTSI satisfying IRE and IEC, so that CS$\indt$ and CL$\indt$ are satisfied. This is the first time
that causal properties have been shown for HO$\pi$.

\subsection{Reversible $\pi$-calculus (R$\pi$)}\label{sec:pi}
We consider the (uncontrolled) reversible semantics for
$\pi$-calculus defined in~\cite{CristescuKV13}. We restrict the
attention to reachable processes. The semantics is an LTS
semantics.
Independence is given as concurrency which is defined for consecutive transitions~\cite[Definition
  4.1]{CristescuKV13}.  CC holds~\cite[Theorem~4.5]{CristescuKV13}.

Our results are not directly applicable to R$\pi$,
since SP holds up to label equivalence of transitions on opposite sides
of the diamond,
rather than equality of labels as in our approach.
We would need to extend axiom SP and the definition of causal equivalence to allow for label equivalence in order to directly handle R$\pi$ using our axiomatic method.

We can however apply our theory to an LTSI obtained by considering labels up-to the equivalence relation $=_\lambda$~\cite[just before Lemma 4.3]{CristescuKV13}, which intuitively avoids to observe when a name is being extruded.
Notice that the Loop Lemma holds in this new LTSI as well.
However, the concurrency relation is given on consecutive transitions, and the same for their SP. Nevertheless, we can define independence as follows: $t \ind_\pi u$ iff $t$ and $u$ are coinitial and $t$ and $\rev u$ are concurrent. Notice that since $t$ and $u$ are coinitial then $t$ and $\rev u$ are consecutive. 
\begin{lemma}
  $\ind_\pi$ is symmetric.
\end{lemma}
\begin{proof}
  We have to show that $t$ and $\rev u$ are concurrent iff $\rev t$ and $u$ are concurrent. Since concurrency is defined as the complement of structural causality and contextual causality~\cite[Definition 4.1]{CristescuKV13}, it is enough to prove that $t$ and $\rev u$ are structural or contextual causal iff $\rev t$ and $u$ are. For structural causality, it follows from the definition~\cite[Definition 4.1]{CristescuKV13}. For contextual causality, it follows from~\cite[Proposition 4.2]{CristescuKV13}.
\end{proof}

With this definition of independence SP holds~\cite[Lemma 4.3]{CristescuKV13}. WF
holds as well since each backward step consumes at least a memory.
BTI has been proved as part of the proof of PL in~\cite[Lemma
  14]{CristescuPhD}.  As a result we obtain a proof of CC much simpler
than the one in~\cite[Theorem 11]{CristescuPhD} (note that causal
equivalence in~\cite[Definition 4.4]{CristescuKV13} is formalised
up-to $=_\lambda$ as well).

Independence is coinitial by construction.
We have to prove PCI and CIRE. Unfortunately, we cannot exploit CLG, since it does not hold, as is clear from the definition of structural cause~\cite[Definition 4.1]{CristescuKV13}, one of the ingredients of the concurrency relation. Thus we need to go for a direct proof. 

\begin{lemma}
  CIRE holds in the LTSI for R$\pi$.
\end{lemma}
\begin{proof}
  Concurrency is defined as the complement of structural causality and
  contextual causality~\cite[Definition
    4.1]{CristescuKV13}. Contextual causality is defined on
  labels~\cite[Proposition 4.2]{CristescuKV13}. Structural causality
  depends on whether the $i$ components of the two labels occur in the
  same memory in a specific relation~\cite[Definition
    2.2]{CristescuKV13}. However, one can notice that $i$ can only
  occur in the memory of one of the threads participating to the
  action (see~\cite[Table 1]{CristescuKV13}), which are the same in
  transitions in the same event. The thesis follows.
\end{proof}
\begin{lemma}
  PCI holds in the LTSI for R$\pi$.
\end{lemma}
\begin{proof}
  Similar to the one above.
\end{proof}
Using PCI and CIRE, we get the CS$\ci$ and CL$\ci$ safety and liveness properties. 
Applying mapping $g$ from Section~\ref{sec:coinitial}, we get a general 
pre-reversible LTSI satisfying IRE and IEC, so that CS$\indt$ and CL$\indt$ are satisfied.
Notice that the notion of independence is not influenced by the abstraction on labels; hence the results can be reflected on the original LTSI of R$\pi$.


\subsection{Reversible Internal $\pi$-calculus with Extrusion Histories ($\pi$IH)}
The calculus $\pi$IH~\cite{GPY21} is based on the work of Hildebrandt \emph{et al.}~\cite{HJN19},
which uses extrusion histories and locations to define a stable non-interleaving early operational semantics for the $\pi$-calculus.
Locations and extrusion histories are used to define independence of actions.
This notion of independence differs from the ones considered in the other case studies in that it allows actions with conflicting causes to be independent.
Despite this major difference, it is shown in~\cite{GPY21} that nearly all our (non-structural) axioms are satisfied
(SP, BTI, WF, PCI and IRE); the only exception is that IEC fails,
because a process can have independent transitions with conflicting causes without having a single state where equivalent transitions can both be performed.
We use IEC to show RPI (Proposition~\ref{prop:RPI}).
However RPI is shown in~\cite{GPY21}  for $\pi$IH without the need for IEC,
using the fact that independence is defined on transition labels.
In fact, LG holds for $\pi$IH, from which we can deduce PCI, IRE and RPI by Proposition~\ref{prop:LG}.
It follows that all the properties listed in Table~\ref{t:list} hold for $\pi$IH, with the exception of IEC, IC and CLG.

\subsection{Reversible Erlang}\label{sec:erlang}
We consider the uncontrolled reversible (reduction) semantics for Erlang
in~\cite{LaneseNPV18}. We restrict our attention to reachable
processes. 
In order to
have more informative labels we can consider the annotations defined
in~\cite[Section 4.1]{LaneseNPV18}. We can then define coinitial transitions to be independent
iff they are concurrent~\cite[Definition 12]{LaneseNPV18}.  

We next discuss the validity of our axioms in reversible Erlang.
SP is proved in~\cite[Lemma 13]{LaneseNPV18} and BTI is trivial from the definition 
of concurrency~\cite[Definition 12]{LaneseNPV18}.  WF holds since the pair of non-negative integers 
(total number of elements in history, total number of messages queued) ordered under
lexicographic order decreases at each backward
step. Intuitively, each step but the ones derived using the rule for reverse sched 
(see~\cite[Figure~11]{LaneseNPV18}) consumes an item of memory, and each step derived using 
rule reverse sched removes a message from a process queue. Finally, PCI and CIRE hold since CLG holds for the LTSI with annotated labels,
and by Proposition~\ref{prop:CLG}.

Since this setting is very similar to the one of HO$\pi$
(both calculi have a reduction semantics and a coinitial notion of independence defined on enriched labels),
we get the same results as for
HO$\pi$ (described in Section~\ref{sec:hopi}), including CC, and causal safety and liveness.

\subsection{Reversible occurrence nets}
We consider occurrence nets, which are the result of unfolding Place/Transition nets, and their reversible versions~\cite{MMU19,MMU20,MelgrattiMPPU2020}. 
Reversible occurrence nets are occurrence nets 
(1-safe and with no backward conflicts)
extended with a backward (reverse in the terminology of~\cite{MMU20}) transition name $\overleftarrow{{\sf t}}$ for each forward transition name ${\sf t}$. We write $t, u$ (note the $italic$ font) for forward or backward transition names, and $\overleftarrow{t}, \overleftarrow{u}$ for their backward or forward duals. We use ``transition name'' to mean forward or backward transition name.
They give rise to an LTS where states 
are pairs $(N,m)$ with $N$ a net and $m$ a marking. A computation that represents firing a (forward or backward)
transition name $t$ in $(N,m)$ and resulting in $(N,m')$ is given by a firing relation $(N,m)\tran{t} (N,m')$~\footnote{
We use ``transition names'' in this subsection to name the members of the set of transitions which, together with the set of places,
are part of the definition of Place/Transition nets or occurrence nets. This distinguishes them from our transitions, which are called firings in Place/Transition nets and occurrence nets.}. 
Independence is the concurrency relation $\co$ which is defined between arbitrary firings as follows:
two firings are concurrent if their transition names are concurrent, that is when they are not in conflict 
and do not cause each other~\cite[Section 3]{MMU19,MMU20}. The last two notions are defined in terms 
of conditions on pre- and postset relations on transition names.
Hence, we get an LTSI with general independence. Note that transition names are unique. 

Properties SP and PL are shown as~\cite[Lemma~4.3]{MMU20} and \cite[Lemma~4.4]{MMU20}, respectively. 
Then CC is proved (over several pages) as~\cite[Theorem~4.6]{MMU20} using SP and PL.
The causal safety and causal liveness properties are not considered in \cite{MMU19,MMU20}.
However, a form of such properties is discussed in~\cite{MelgrattiMPPU2020} in the setting of reversible prime event structures; we discuss this point in Section~\ref{sec:related}.

We can obtain 
causal safety and causal liveness properties, as well as PL and CC, for reversible occurrence nets using our axiomatic approach.
The following lemma will be helpful.
\begin{lemma}\label{lem:nocausation}
Let $t$ and $u$ be enabled and coinitial (forward or backward) transition names. Then $t$ does not cause $u$. If additionally  $t$ and $u$ are backward, 
then they are not in conflict.
\end{lemma}
\begin{proof}
 Assume for contradiction that $t$ causes $u$. 
So there is a place, say $a$, in the preset of  $u$ such that $t$ causes $a$. Since $u$ is enabled there is a token in $a$. 
Also, since $t$ is enabled, after it fires a second token will arrive in $a$, thus
contradicting the 1-safe property of occurrence nets.

Let $t$ and $u$  be $\overleftarrow{{\sf t}}$ and 
  $\overleftarrow{{\sf u}}$ respectively.  Assume for contradiction that  they are in conflict. This means that they share a place, say $a$, in their presets. Hence, ${\sf t}$ and ${\sf u}$ share $a$ in their postsets, which contradicts the no backwards conflict property of occurrence nets. 
\end{proof}

We can now combine Lemma~\ref{lem:nocausation} with the conditions in~\cite[Lemma~3.3]{MMU20} of when enabled and coinitial $t$ and $u$ are concurrent.

\begin{lemma}\label{lem:on-concurrency}
Let $t$ and $u$ be enabled and coinitial (forward or backward) transition names. Then
$t\co u$ iff $t$ and $u$ are backward or they are not in an immediate conflict.
\end{lemma}
As a consequence, BTI holds.

\begin{lemma}
  BTI holds in the LTSI for reversible occurrence nets.
\end{lemma}

WF holds because there are no forward cycles of firings in occurrence nets, hence 
no infinite reverse paths. This gives us PL and CC.
Next, we prove PCI. 
\begin{lemma}
PCI holds in the LTSI for reversible occurrence nets.
\end{lemma}
\begin{proof}
Consider enabled coinitial  firings $\phi_1, \phi_2$ with transition names $t, u$ respectively, and assume $\phi_1\co \phi_2$. Hence $t\co u$. We get a commuting diamond by SP, where the opposite sides have the same transition names. Since $t\co u$, we have  $\overleftarrow{t}\co u$ by~\cite[Lemma 3.4]{MMU20}, so PCI holds. 
\end{proof}
This gives us a pre-reversible LTSI, and thus CS$\ci$ and CS$_<$ hold.

Given a pair of enabled coinitial concurrent  transition names we get a commuting diamond by SP, and the pairs of  coinitial transition names in all corners of the diamond are concurrent. Events can then be defined on firings in such diamonds as in Definition~\ref{def:sqeqt}, and we can show IRE.
\begin{lemma}\label{lem:IRE-occnets}
  IRE holds in the LTSI for reversible occurrence nets.
\end{lemma}
\begin{proof}
Let $\phi_1, \phi_2$ be firings with $t, u$ respectively, and let $\phi_1\co \phi_2$. 
This means that $t\co u$. Since any $\phi_1'$ equivalent to $\phi$ has the same transition name $t$, $t\co u$ gives us
$\phi_1'\co \phi_2$.
\end{proof}

Since IRE implies CIRE we obtain CL$\ci$ (or CL$_<$). We also have CS$\indt$ and CL$\indt$ as IRE holds.

An alternative proof strategy would be to show CLG first, but we believe this approach leads to more complex technicalities, and we would still need to prove IRE, hence we have preferred the approach above.
\subsection{Reversible sequential systems}
In \emph{sequential systems} there is no concurrency.
Hence, in this section, we represent them as LTSIs where the independence
relation, modelling concurrency, is empty.
This is for instance the case for Janus programs~\cite{YokoyamaG07} or CCSK processes without parallel composition such as the ones studied in~\cite{BR23}.
In this setting,
SP, PCI, IRE and IEC hold trivially. 
Moreover, BTI is equivalent to backward determinism, which is
the main condition required for reversibility in a sequential setting (see, e.g., Janus~\cite{YokoyamaG07}).

\begin{definition}[Backward determinism]
An LTSI is backward deterministic iff $P \ftran{a} Q$ and $P'
\ftran{a'} Q$ imply $P = P'$ and $a=a'$.
\end{definition}

\begin{proposition}\label{prop:sequential}
A sequential system satisfies BTI iff it is backward deterministic.
\end{proposition}
\begin{proof}
For the left to right implication, assume towards a contradiction that
the system satisfies BTI but it is not backward deterministic. Then
there are $P \ftran{a} Q$ and $P' \ftran{a'} Q$ with $P \neq P'$ or $a
\neq a'$. By the Loop Lemma we have the reverse transitions, which are
coinitial and backwards, hence by BTI they need to be independent,
what is a contradiction since the independence relation is empty.

For the right to left implication, take two backward coinitial
transitions $t,t'$. By applying the Loop Lemma there exist $\rev t,
\rev t'$. One can notice that $\rev t,\rev t'$ satisfy the hypothesis
of backward determinism. Hence, $\rev t=\rev t'$ and $t=t'$. Hence BTI
trivially holds.
\end{proof}

WF does not hold in general and needs to be assumed.

If we assume WF then all our results hold, but they all become trivial
or almost trivial.  E.g., all events are singletons. Also, all the notions of causal liveness coincide, and they state that the last
transition can always be undone, but this is just one direction of the
Loop Lemma. Similarly, all the notions of causal safety do coincide, and they require that only the last transition can
be undone.

\section{Related Work}\label{sec:related}
Causal Consistency (CC), Parabolic Lemma (PL) and informal versions of Causal Safety and Liveness (CS, CL), 
the main general properties of reversible computation considered in this paper, were
proposed by Danos and Krivine~\cite{DK04}. Since then, many reversible process calculi or formalisms 
have been developed as we have described in the Introduction. Most of them 
use memories to save information lost when computing forwards, which can be easily retrieved 
when computing in reverse. Concurrency relation between coinitial transitions is typically defined 
in terms of structural conditions on the
memories of the transitions. In order to show that reversibility 
is well-behaved, PL and then CC is proved. In contrast, CS and CL (in any of the variants we considered), or properties close to them, 
have not been widely considered.

Information needed for undoing of computation in a process calculus can be saved differently.  An alternative
method was proposed for reversing a process calculus given by
a general format of SOS rules in~\cite{PU06,PU07}.
When applied to CCS it produces CCSK, where reversible processes maintain 
their syntax as they compute, and executed actions are marked with \emph{communication keys}. 
When computation reverses, keys are removed, thus returning processes to their original form. 
This approach has a drawback in that it is not easy 
to define a concurrency relation purely on transition labels. As a result, proving CC in 
the traditional way is not straightforward. Hence, slightly different properties are proved 
to show that the resulting reversible calculi are
well-behaved. The main property is Reverse Diamond (RD):
if $Q \tran a P$, $R \tran b P$ and $Q \neq R$, then there is $S$ such that $S \tran a R$ and
$S \tran b Q$. In our setting, RD can be proved from the Loop Lemma, BTI and SP. It is worth noting that PL can be shown for CCSK mainly using RD~\cite[Lemma 5.9]{PU07}.
Moreover, a form of CC for forward computation is shown~\cite[Proposition 5.15]{PU07}: 
two forward computations from the same start to the same 
endpoint are \emph{homotopic}~\cite{vG96}, meaning that one computation can be transformed into the other by swapping 
adjacent transitions in commuting diamonds. In effect, concurrency is represented as commuting diamonds in 
the LTSs for reversible calculi obtained by applying the method in~\cite{PU06,PU07}.

A more abstract approach to defining desirable properties for reversibility was taken
in~\cite{PU07a}. General LTSs were considered instead of LTSs for specific reversible calculi, and two 
sets of axioms were proposed. The first set inherited RD and Forward Diamond (FD) from \cite {PU06,PU07},
and also included WF, UT and Event Determinism (ED)~\cite{SNW96,vG96}:
if $P \tran a Q$ and $P \tran a R$, and $(P,a,Q) \sqeqt (P,a,R)$, then $Q = R$.
ED is not a consequence of our basic axioms. Consider the LTS~\cite[Fig. 1]{PU07a}, and add coinitial independence 
using BTI and PCI.  The resulting LTSI is pre-reversible and satisfies CLG, yet it fails ED. 
LTSs satisfying the
five axioms above are called \emph{prime} LTSs and are shown
to correspond to prime event structures.
Several interesting properties were proved for prime LTSs, including
RED (event determinism for backward transitions, which follows from BLD in our setting) and
NRE which we also consider here. 
The second set of axioms aimed at providing local versions of FD, ED and RED. 



As we have mentioned in the Introduction, a combined causal safety and liveness property
has been formulated in~\cite[Corollary 22]{LaneseNPV18}.
A form of causal safety and liveness properties has been defined in the setting 
of reversible event structures in~\cite{PU13,PU15}.
A reversible event structure is called \emph{cause-respecting}
if an event cannot be reversed until all events it has caused have also been reversed, and 
it is \emph{causal} if it is cause-respecting and a reversible event can be reversed
if all events it has caused have been reversed~\cite[Definition~3.34]{PU15}. Causal reversible prime event structures
are considered in~\cite{MelgrattiMPPU2020} as well, where it is shown that they correspond precisely
to reversible occurrence nets.

Another related work is~\cite{DanosKS07},
which like ours takes an abstract view, though based on
category theory. However, its results concern irreversible actions,
and do not provide insights in our setting, where all actions are
reversible. The only other work which takes a general perspective
is~\cite{BernadetL16}, which concentrates on how to derive a
reversible extension of a given formalism. However, proofs concern a
limited number of properties (essentially our CC), and hold only 
for extensions built using the technique proposed there. 
An approach similar to that in~\cite{PU07,BernadetL16} is taken in~\cite{LaneseM20}, which focuses on
systems modelled using reduction semantics. In order to prove
properties of the reversible systems they build they use our theory
(taken from the conference version of the present
paper~\cite{LanesePU20}), hence this can be taken as an additional
case study for our results.
Finally,~\cite{EKM19} presents a number of properties such as, for example,
backward confluence, which arise in the context of reversing
of multiple transitions at the same time (called a step) in Place/Transition nets.

\section{Conclusion and Future Work}\label{sec:conclusion}


The literature on causal-consistent reversibility (see, for example the
early survey~\cite{LMT14}) has a number of proofs of results
such as PL and CC, all of which are instantiated to a specific calculus, language or formalism.  
We have taken here a complementary and more general approach, analysing the properties of
interest in an abstract and language-independent setting. In particular,
we have shown how to prove the most relevant of these properties from a
small number of axioms. Among the properties, we discussed in detail the formalisation of Causal Safety and Causal Liveness, which were mostly informally discussed in the literature.

The approach proposed in this paper opens a number of new
possibilities. Firstly, when devising a new reversible formalism, our
results provide a rich toolbox to prove (or disprove) relevant
properties in a simple way. Indeed, proving the axioms is usually much simpler than proving the properties directly. This is particularly relevant since
causal-consistent reversibility is getting applied to more and more
complex languages, such as Erlang~\cite{LaneseNPV18}, where direct
proofs become cumbersome and error-prone.
Secondly, our abstract proofs are relatively easy to formalise in a
proof-assistant, which is even more relevant given that this will
certify the correctness of the results for many possible instances.
Another possible extension of our work concerns integrating into our framework mechanisms to control reversibility~\cite{LaneseMS12}, such as
 a rollback operator~\cite{LMSS11} or irreversible actions~\cite{DK05}. For the latter we could take
inspiration from the above-mentioned~\cite{DanosKS07}.

\section*{Acknowledgements}
  This work has been partially supported by COST Action IC1405 on Reversible Computation - Extending Horizons of Computing. The first author has also been partially supported by the French ANR project DCore ANR-18-CE25-0007 and by the INdAM-GNCS project CUP\_E55F22000270001 \emph{Propriet\`a Qualitative e Quantitative di Sistemi Reversibili}.
  The third author has been partially supported by the JSPS Invitation Fellowship S21050.

\bibliographystyle{plain}
\bibliography{axrev}

\providecommand{\url}[1]{\texttt{#1}}
\providecommand{\urlprefix}{URL }
\providecommand{\doi}[1]{https://doi.org/#1}


\end{document}